\documentclass[a4paper,11pt]{article}
\usepackage{jheppub} % for details on the use of the package, please see the JINST-author-manual
\usepackage{lineno}
\usepackage{bm}
\newcommand{\muEW}{\mu_{\text{\tiny EW}}}
\newcommand{\Eq}[1]{Eq.~(\ref{#1})}
\newcommand{\Tab}[1]{Table~(\ref{#1})}
\newcommand{\Fig}[1]{Fig.~(\ref{#1})}
\newcommand{\Sec}[1]{Sec.~(\ref{#1})}
\newcommand{\App}[1]{App.~\ref{#1}}
\newcommand{\EqRange}[2]{Eqs.~(\ref{#1})-(\ref{#2})}

\newcommand{\cQL}{{\bm c}_{Q_L}}
\newcommand{\cuR}{{\bm c}_{u_R}}
\newcommand{\cdR}{{\bm c}_{d_R}}

\newcommand{\MKpia}{\widetilde{\mathcal{M}}}

\title{
Probing CP and flavor violation in neutral kaon decays with ALPs
}
\author[a]{Reuven Balkin,}
\author[a]{Stefania Gori,}
\author[b,c]{Christiane Scherb}
\affiliation[a]{Department of Physics, University of California Santa Cruz and Santa Cruz Institute for Particle Physics, 1156 High St., Santa Cruz, CA 95064, USA}
\affiliation[b]{Berkeley Center for Theoretical Physics, 
Department of Physics,
University of California, Berkeley, CA 94720, USA}
\affiliation[c]{Theoretical Physics Group, Lawrence Berkeley National Laboratory, Berkeley, CA 94720, USA}
\emailAdd{rebalkin@ucsc.edu}
\emailAdd{sgori@ucsc.edu}
\emailAdd{cscherb@lbl.gov}

\abstract{
We analyze the three-body decays of the long-lived neutral kaon $K_L \to \pi\pi a$, where $a$ is an axion-like particle (ALP), and compare them to the two-body decay $K_L \to \pi^0 a$. 
While the latter requires both flavor violation (FV) and $CP$ violation (CPV), the former can proceed via FV alone, allowing the ratio of decay rates to serve as a probe of CPV of the underlying UV theory. We emphasize the importance of weak-interaction-induced contributions, often neglected in recent calculations.
We explore both minimal and non-minimal flavor-violating scenarios, and identify classes of models where ALP production from neutral three-body decays is comparable to—or even dominates over—the two-body decay, despite its reduced phase space.
Finally, we discuss the phenomenological implications of our results and show how these decays can provide complementary probes of ALP couplings beyond those accessible via charged kaon channels.
}

\begin{document}
\maketitle
\flushbottom
%%%%%%%%%%%%%%%%%%%%%%%%%%%%%%%%%%%%%%
%%%%%%%%%%%%%%%%%%%%%%%%%%%%%%%%%%%%%%
\section{Introduction}
\label{sec:intro}
%%%%%%%%%%%%%%%%%%%%%%%%%%%%%%%%%%%%%%
%%%%%%%%%%%%%%%%%%%%%%%%%%%%%%%%%%%%%%
The approximate $U(3)^3$ quark flavor symmetry of the Standard Model (SM), broken explicitly only by the quark Yukawa interactions, is a well-tested feature of Nature.
This symmetry explains several observed phenomena, notably the suppression of flavor-changing neutral-current (FCNC) processes.
Simple parameter counting reveals that the quark sector of the SM contains only a single physical $CP$-violating phase\footnote{More accurately, the CKM phase is the only source of $CP$-violation which appears in the Standard Model at the perturbative level.
The QCD $\theta$ angle is another $CP$-violating parameter, which however appear only from non-perturbative effects.}, making it the sole experimentally observed source of $CP$ violation~\cite{Christenson:1964fg}.
The phase appears in the Cabibbo-Kobayashi–Maskawa (CKM) matrix, which is responsible for flavor-changing processes in the SM.
Thus, flavor-violation (FV) and $CP$-violation (CPV) go hand-in-hand within the SM.

Since generically FCNC and CPV processes are highly suppressed in the SM, they offer sensitive probes of physics beyond the Standard Model (BSM).
Such processes are sensitive even to small contributions from BSM physics, which are typically suppressed by a high UV scale $\Lambda$. 
If the new physics (NP) is heavy ($m_{\text{\tiny NP}} \sim \Lambda$), its contribution can be observed indirectly and studied systematically via an effective field theory (EFT) approach.
If, instead, the new physics is light ($m_{\text{\tiny NP}} \ll \Lambda$), such particles can be searched for more directly — e.g., via the production in flavor-violating hadron decays, or at the LHC.

In this work we focus on axion-like particles (ALPs), whose small mass can naturally arise from an approximate shift symmetry.
ALPs have seen a revival of theoretical and experimental interest in recent years.
The theoretical motivation for the existence of these particles is broad. They originally appeared in SM extensions which address the strong CP problem~\cite{Peccei:1977hh,Peccei:1977ur,Weinberg:1977ma,Wilczek:1977pj}, are compelling dark matter candidates~\cite{Abbott:1982af,Dine:1982ah,Preskill:1982cy,Marsh:2015xka,Adams:2022pbo} and emerge generically from string theory frameworks~\cite{Arvanitaki:2009fg}. 
The experimental effort is equally broad, spanning searches relying on their primordial abundance, their effect on cosmology and astrophysical systems, or their production at accelerator experiments.
For a recent review, see~\cite{PDG:axions}.

Unlike charged kaon decays,  neutral kaon decays provide a unique probe of both the FV \emph{and} the CPV properties of the ALP couplings to the SM.
New flavored couplings could introduce new sources of CP violation beyond the CKM phase. 
Conversely, if no new sources of FV are introduced, the theory is said to be minimally flavor-violating (MFV)~\cite{DAmbrosio:2002vsn}.
As we will discuss, since the ALP couplings are hermitian in flavor space, the leading terms in the MFV spurion expansion do not introduce new CPV phases.
Thus, observing neutral kaon decays involving ALPs may provide some hints regarding both the underlying CP and flavor structure of Nature.

Two- and three-body neutral kaon decays can proceed either through \emph{direct} flavor-changing couplings or \emph{indirectly} via SM flavor-changing weak interactions combined with flavor-preserving ALP couplings.
In this work, we emphasize the importance of the indirect weak-interaction contributions,  which were studied for the two-body decays~\cite{Bauer:2021wjo}, but are often not discussed in analyses of the three-body decays~\cite{MartinCamalich:2020dfe,Cavan-Piton:2024pqp,DiLuzio:2023cuk}.
To this end, we adopt the approach of~\cite{Bauer:2021wjo} to ensure the results are basis-independent, i.e. depend only on physical combinations of couplings. 
Unlike for the two-body decays, our calculation for three-body decays requires the inclusion of naively factorizable contributions, which cancel out non-physical contributions originating from contact terms.
The flavor-changing coupling responsible for directly mediating the decay arises either at tree-level or at one-loop~\cite{Bauer:2020jbp}.

We focus on the decays of the approximately $CP$-odd eigenstate $K_L$ to pions and an ALP,
in particular on the less-studied three-body decays $K_L \to \pi \pi a$, which require FV but not CPV,
and compare them to the two-body decay $K_L \to \pi^0  a$, which requires both FV \emph{and} CPV.
Thus, the ratio of these rates $\Gamma(K_L \to \pi \pi a)/\Gamma(K_L \to \pi^0  a)$ provides a probe of CPV of the BSM sector. 
In non-MFV scenarios with sizable new sources of FV, the rate ratio depends on whether $CP$ is also violated by the ALP couplings.
For MFV theories, the situation is more nuanced, depending on which coupling mediates the process.
Several hierarchies of rates are then possible depending on which couplings are present in the UV theory.
In particular, we find that in some theories the three-body decay can dominate over the two-body decay rate despite the reduced phase-space volume.

The paper is organized as follows. 
We start by presenting the ALP EFT at the UV scale in \Sec{sec:coupling_and_REG} and map it to the IR theory just above the QCD scale, taking into account the running effects in off-diagonal couplings.
Next, we present the relevant amplitudes of kaon decays calculated at leading order in $\chi$PT in \Sec{sec:decay_rates}, with additional details on the calculation provided in \App{app:chi_PT}.
Following a short discussion of the structure of the FV coupling in \Sec{sec:FV_couplings}, we proceed analyzing the ratio of rates in \Sec{sec:rates} and identify the resulting rate hierarchies under different coupling assumptions, with additional details given in \App{app:PSI_and_ratios}, \App{sec:non_MFV}, and \App{app:MFV_details}.
We discuss the phenomenological implications using one benchmark model in \Sec{sec:pheno} before summarizing and concluding in \Sec{sec:conc}.  
Details on the recast of experimental limits used in \Sec{sec:pheno} are given in \App{app:exp_recasts}.
%%%%%%%%%%%%%%%%%%%%%%%%%%%%%%%%%%%%%%
%%%%%%%%%%%%%%%%%%%%%%%%%%%%%%%%%%%%%%
\section{ALP effective theory}
\label{sec:coupling_and_REG}
%%%%%%%%%%%%%%%%%%%%%%%%%%%%%%%%%%%%%%
%%%%%%%%%%%%%%%%%%%%%%%%%%%%%%%%%%%%%%
Our starting point is a theory at a UV scale $\Lambda$ above the electroweak scale $\muEW$. 
We consider a pseudoscalar, $a$, with approximately shift-symmetric couplings  to the SM fields. 
The most general EFT up to dimension-5 operators is given by
\begin{align}
\label{eq:UV_Lagrangian}
    \mathcal{L}_a(\Lambda) = &\frac12 (\partial_\mu a)( \partial^\mu a)-\frac12 m_a^2\,a^2+\frac{\partial_\mu a}{f}\sum_F \bar{F}\, {\bm c}_F \,\gamma_\mu F  \nonumber
    \\
    &+c_{GG} \frac{\alpha_s}{4\pi} \frac{a}{f} G_{\mu\nu}^a \tilde{G}^{a,\mu\nu}
    +
    c_{WW} \frac{\alpha_2}{4\pi} \frac{a}{f} W_{\mu\nu}^i \tilde{W}^{i,\mu\nu}
    +
    c_{BB} \frac{\alpha_1}{4\pi} \frac{a}{f} B_{\mu\nu} \tilde{B}^{\mu\nu}\,.
\end{align}
The shift symmetry of $a$ is explicitly broken by its mass term $m_a^2$ and by non-perturbative QCD effects at low energies. 
The quark and lepton couplings run over all SM chiral multiples $F\in\{Q_L,u_R,d_R,L_L,e_R\}$, where ${\bm c}_F$ are hermitian matrices in flavor space. 
The operator $\mathcal{O}_H \equiv (\partial_\mu a)(H^\dagger \overset{\leftrightarrow}{\partial^\mu}H) $ could also be added to the Lagrangian, but is redundant and can be eliminated by a one-parameter family of  field redefinitions~\cite{Bauer:2020jbp}. 
Next, we consider the low-energy theory at a scale $\mu<\muEW$, just above the QCD confinement scale:
\begin{align}
  \mathcal{L}_a(\mu)  = &\frac12 (\partial_\mu a)^2-\frac12 m_a^2\,a^2
+c_{GG} \frac{\alpha_s}{4\pi} \frac{a}{f} G_{\mu\nu}^a \tilde{G}^{a,\mu\nu}+c_{\gamma\gamma} \frac{\alpha}{4\pi} \frac{a}{f} F_{\mu\nu} \tilde{F}^{\mu\nu}\nonumber
\\
&+\frac{\partial^\mu a}{f}(\bar{q}_L {\bf k}_Q \gamma_\mu q_L+\bar{q}_R {\bf k}_q \gamma_\mu q_R)  \,,
\label{eq:UV_matching_theory}
\end{align}
where $q=\{u,d,s\}$ denotes the light quark fields, and couplings to leptons have been dropped. 
 The coupling to photons is given by $c_{\gamma\gamma}=c_{WW}+c_{BB}$, where we neglected the $\mathcal{O}(m_a^2/m_t^2,m_a^2/m_W^2)$ loop contributions from weak-scale particles.\footnote{The effective and basis-independent coupling to photons does include contributions from light degrees of freedom as well, see \App{app:photon_coupling} for more details.}
Without loss of generality, we choose the basis in \Eq{eq:UV_Lagrangian} in which the up-type SM Yukawa matrix is diagonal.
The mass basis for down-type quarks is then obtained by the rotation $d_L \to {\bm V} d_L$, where 
$\bm V$ is the CKM matrix.
We identify
\begin{align}
    {\bm k}_q = \begin{pmatrix}
    [{\bm c}_{u_R}]_{11} & 0 & 0\\
     0& [{\bm c}_{d_R}]_{11} & 
     \kappa_R
     \\
     0& 
          \kappa^\dagger_R
     & [{\bm c}_{d_R}]_{22} 
    \end{pmatrix}\,, \;\;
        {\bm k}_Q = \begin{pmatrix}
    [{{\bm c}}_{Q_L}]_{11} & 0 & 0\\
     0& [\hat{{\bm c}}_{Q_L}]_{11} & 
          \kappa_L
     \\
     0& 
          \kappa^\dagger_L
     & [\hat{{\bm c}}_{Q_L}]_{22} 
    \end{pmatrix}\,,
\end{align}
where $\hat{{\bm c}}_{Q_L} \equiv {\bm V}^{\dagger}{\bm c}_{Q_L}{\bm V}$. 
We now focus on the flavor-violating couplings that mediate $s\to d$ transitions, defining the corresponding vector and axial combinations $\kappa_{V},\kappa_{A}$ as,
\begin{align}
    \mathcal{L}_{\text{\tiny FV}}=& \frac{\partial_\mu a}{f} (\kappa_R \bar{d}_R  \gamma_\mu s_R+ \kappa_L \bar{d}_L  \gamma_\mu s_L + \text{h.c})
\equiv \frac{\partial_\mu a}{2f} (\kappa_V \bar{d}  \gamma_\mu s+ \kappa_A \bar{d}  \gamma_\mu \gamma_5 s + \text{h.c})\,.
\end{align}
At leading order, the off-diagonal coupling $\kappa_{R}$ does not run below the UV scale $\Lambda$~\cite{Bauer:2020jbp} and is given by, 
\begin{align}
    \kappa_R = [{\bm c}_{d_R}(\Lambda)  \big]_{12} \,.
\end{align}
The coupling to the left-handed quarks receives radiative corrections and is given by
 \begin{align}\label{eq:KL}
    \kappa_L &= \big[\hat{{\bm c}}_{Q_L}\big]_{12}+{V}^*_{td}{V}_{ts}\Delta \kappa_L\,,
\end{align}
The first term is a direct consequence of the UV couplings and the transition to the mass basis,
\begin{align}
\label{eq:UV_off_diag_couplings}
     \big[\hat{{\bm c}}_{Q_L}\big]_{12}  &= V^*_{qd}V_{q's}  \big[ \cQL\big]_{q q'} 
     =[\cQL(\Lambda)  \big]_{12} \nonumber
    \\
    & +  {V}^*_{cd}{V}_{cs}\left(\big[  \cQL(\Lambda)  \big]_{22}-\big[  \cQL(\Lambda)  \big]_{11}\right) \nonumber
    \\
    & + {V}^*_{td}{V}_{ts}\left(\big[  \cQL(\Lambda)  \big]_{33}-\big[  \cQL(\Lambda)  \big]_{11}\right) +...\,,
\end{align}
where we omitted terms suppressed by smaller CKM matrix elements.
The radiative correction is given by~\cite{Bauer:2020jbp}
\begin{align}
\label{eq:rad_correction}
\Delta \kappa_L = n_{t}c_{tt}(\Lambda)+n_{G}\tilde{c}_{GG}(\Lambda)+n_{W}\tilde{c}_{WW}(\Lambda)+n_{B}\tilde{c}_{BB}(\Lambda)\,,
\end{align}
where we denote
\begin{align}
\label{eq:nt_def_1}
    n_t(\muEW,\Lambda) &\equiv   \frac{1}{2} \frac{ \alpha_t( \muEW) }{ \alpha_s(\muEW )}\left(1-\left[{\frac{\alpha_s(\Lambda )}{\alpha_s(\muEW )}}\right]^{1/7}\right)\,,
    \\
    \label{eq:nG}
    n_{G}(\Lambda) &\equiv \frac{2}{\pi^2}\int_\Lambda^{\muEW}\frac{d\mu}{\mu}
    n_t(\muEW,\mu)
     \alpha_s^2(\mu)\,,
     \\
\label{eq:nt_def_2}
    n_{W}(\Lambda) &\equiv \frac{9}{16\pi^2}\int_\Lambda^{\muEW}\frac{d\mu}{\mu}
    n_t(\muEW,\mu)
     \alpha_2^2(\mu)- \frac{3\alpha_t(\muEW)}{8\pi^2} \frac{\alpha_2}{ s_w^2}\,
    \frac{1-x_t+x_t\ln x_t}{\left(1-x_t\right)^2}\,,
                   \\
    n_{B}(\Lambda) &\equiv \frac{17}{48\pi^2}\int_\Lambda^{\muEW}\frac{d\mu}{\mu}
    n_t(\muEW,\mu)
     \alpha_1^2(\mu)\,,
     \label{eq:nt_def_4}
\end{align}
with $\alpha_t \equiv y_t^2/4\pi$ and $x_t \equiv m_t^2/m_W^2$.
Since the off-diagonal coupling $\kappa_{L}$ does not run below the electroweak scale $\muEW$, the expressions in \EqRange{eq:nt_def_1}{eq:nt_def_4} do not depend on the value of $\mu$.
The UV couplings appearing in \Eq{eq:rad_correction} are defined in terms of the couplings in \Eq{eq:UV_Lagrangian} as follows,
\begin{align}
\label{eq:coef_def_1}
    c_{tt}(\Lambda) &\equiv \big[{\bm c}_{u_R}\big]_{33}-\big[\cQL\big]_{33}\,,
    \\
    \tilde{c}_{GG}(\Lambda) &\equiv c_{GG}+\frac12 \text{Tr}\big[{\bm c}_{u_R}+{\bm c}_{d_R}-2 \cQL\big] \,,
        \\
    \tilde{c}_{WW}(\Lambda) &\equiv c_{WW}-\frac12 \text{Tr}\big[3\cQL+{\bm c}_{L_L}\big]\, 
           \\
           \label{eq:coef_def_4}
    \tilde{c}_{BB}(\Lambda) &\equiv c_{BB}+ \text{Tr}\left[\frac43 {\bm c}_{u_R}+\frac13 {\bm c}_{d_R}-\frac16\cQL-\frac12{\bm c}_{L_L}+{\bm c}_{e_R}\right]\,,
\end{align}
where all the couplings on the RHS are evaluated at the UV scale $\Lambda$.
The coupling combinations appearing in \Eq{eq:UV_off_diag_couplings} and \EqRange{eq:coef_def_1}{eq:coef_def_4} are the physical combinations of couplings which are independent of the arbitrary field redefinition used to remove $\mathcal{O}_H$~\cite{Bauer:2020jbp,Bauer:2021mvw}.
In the following section, we shall also encounter the flavor-preserving couplings, e.g.
\begin{align}
    \mathcal{L}_a(\mu) \supset& \frac{\partial_\mu a}{f} (\big[\cdR\big]_{11} \bar{d}_R  \gamma_\mu d_R+ (\big[\hat{{\bm c}}_{Q_L}\big]_{11}\bar{d}_L  \gamma_\mu d_L )+...
\end{align}
and similarly for the up and strange quarks.
These couplings will lead to flavor-changing meson decays mediated by the SM weak interaction.
As opposed to the off-diagonal couplings, the diagonal couplings also run below the weak scale.
However, since these couplings already appear with additional suppression for flavor-violating processes mediated by the SM, we neglect the small RGE effects and CKM-suppressed terms, 
\begin{align}
   &\big[\hat{{\bm c}}_{F}(\mu)\big]_{ii} \approx \big[{\bm c}_{F}(\mu)\big]_{ii} \approx  \big[{\bm c}_{F}(\Lambda)\big]_{ii} \;\;\;\; \text{for $F=\{u_R,d_R,Q_L\}$}\,.
\end{align}
%%%%%%%%%%%%%%%%%%%%%%%%%%%%%%%%%%%%%%
%%%%%%%%%%%%%%%%%%%%%%%%%%%%%%%%%%%%%%
\section{Kaon decay amplitudes}
\label{sec:decay_rates}
%%%%%%%%%%%%%%%%%%%%%%%%%%%%%%%%%%%%%%
%%%%%%%%%%%%%%%%%%%%%%%%%%%%%%%%%%%%%%
\subsection{\texorpdfstring{Kaons and ALPs in ${\bf\chi}$PT}{Kaons and ALPs in chiPT}}
\label{Sec:chipT}
%%%%%%%%%%%%%%%%%%%%%%%%%%%%%%%%%%%%%%
%%%%%%%%%%%%%%%%%%%%%%%%%%%%%%%%%%%%%%
In order to calculate the various decay rates of the kaons, we match the theory above the QCD scale in \Eq{eq:UV_matching_theory} to chiral perturbation theory.
We perform the calculation in a generic basis as outlined in~\cite{Bauer:2021wjo} to ensure the final results depend only on physical combination of parameters, namely combinations that are independent of field redefinition\footnote{
All the results of \Sec{sec:decay_rates} are available as a Mathematica notebook~\cite{githublink}.}.
In this subsection, we report the most important aspects and define the notation. For more details, see \App{app:chi_PT}.

The building block of the chiral Lagrangian is  $\Sigma\equiv e^{2i \Pi/f_\pi}$, with
\begin{align}
\label{eq:PI_def}
\Pi = 
\begin{pmatrix}
\dfrac{\eta_8}{\sqrt{6}} + \dfrac{\pi^{0}}{\sqrt{2}} & \pi^{+} & K^{+} \\[8pt]
\pi^{-} & \dfrac{\eta_8}{\sqrt{6}} - \dfrac{\pi^{0}}{\sqrt{2}} & K^{0} \\[8pt]
K^{-} & \bar{K}^{0} & -\sqrt{\dfrac{2}{3}}\,\eta_8
\end{pmatrix}\,
\end{align}
 parameterizing the mesons as pseudo Nambu-Goldstone fields, and $f_\pi=130\,\text{MeV}$ is the pion decay constant. 
The weak interaction of the mesons is parametrized by the octet operator\footnote{In principle, one should also add the 27-operators~\cite{Bauer:2020jbp,Bauer:2021mvw}, but these are generically negligible compared to the octet. However, see Sec. \ref{sec:GN_bound} for an exception. 
Ref. \cite{Cornella:2023kjq} shows that two additional octet operators  could contribute to processes involving ALPs. However, these operators do not contribute to any SM processes, and therefore their coefficients are unknown. For this work, we assume these coefficients vanish and leave the calculation of their contribution to future work.}
\begin{equation}
   \mathcal{L}_{\text{\tiny weak}} = \frac{4 N_8}{f_\pi^2}[L_\mu L^\mu]^{32}+\text{h.c}\,,\label{eq::Lweak}
\end{equation}
where we defined
\begin{align}
 L_\mu^{ji} &\equiv -\frac{i f^2_\pi} {4}e^{i(\phi_{q_i}^--\phi_{q_j}^-)a(x)/f}[\Sigma (D_\mu \Sigma)^\dagger]^{ji}\,,
 \\
 D_\mu \Sigma &\equiv \partial_\mu \Sigma-i \frac{\partial_\mu a}{f} (\hat{\bf k}_Q\Sigma-\Sigma \hat{\bf k}_q)\,.
 \end{align}
The couplings to quarks, $\hat{\bf k}_Q$ and $\hat{\bf k}_q$, are written after the generic field redefinition
 \begin{align}
q(x) \to \text{exp}\left[-i(\boldsymbol{\delta}_q + \boldsymbol{\kappa}_q \gamma_5 )c_{GG} \frac{a(x)}{f} \right]q(x)\,,
\label{eq:generic_basis}
\end{align}
where $\boldsymbol{\delta}_q$ and $\boldsymbol{\kappa}_q$ are arbitrary diagonal matrices in flavor space. In this basis,
 \begin{align}
&{\bf k}_Q \to \hat{\bf k}_Q(a) = U_-({\bf k}_Q+\phi_q^-)U^\dagger_-\,,
\\
&{\bf k}_q \to \hat{\bf k}_q(a) = U_+({\bf k}_q+\phi_q^+)U^\dagger_+\,,
\end{align}
where $\phi_q^\pm \equiv c_{GG} (\boldsymbol{\delta}_q \pm \boldsymbol{\kappa}_q )$ and $U_\pm(a) \equiv e^{i \phi_q^\pm a/f}$. 

The coefficient in front of the weak operator in Eq. (\ref{eq::Lweak}) is measured and given by
\begin{equation}\label{eq:N8}
   N_8 \equiv -\frac{G_F}{\sqrt{2}} V_{us}V^*_{ud}g_8 f_\pi^2\equiv -|N_8|e^{i\delta_8}\,, \;\;\; |N_8| \approx 1.53\times10^{-7}\,, 
\end{equation}
with $G_F$ the Fermi constant, $G_F=1.1663787\times 10^{-5}$ GeV$^{-2}$ and $|g_8|\sim 5$.
Its strong-interaction phase, $\delta_8$, is not well-known.
The large $N_C$ limit, as well as explicit calculations, seem to point to a small phase~\cite{Cirigliano:2011ny,Gamiz:2003pi,Bijnens:2000im}
\begin{align}
    \delta_8 \approx \text{Im}
\left[ -\frac{V_{td}V^*_{ts}}{V_{ud}V^*_{us}}\right]\approx -6 \times 10^{-4}.
\end{align}
In our numerical calculations, we neglect this phase which is smaller than  other CPV parameters that enter the rates.
Another CPV parameter entering our expressions is the $\varepsilon$  parameter of the kaon-antikaon system.
The kaon mass eigenstates, $K_L$ and $K_S$, are not exact CP eigenstates~\cite{ParticleDataGroup:2020ssz}, and are given by the following superposition of a kaon and an antikaon
\begin{align}
  K_L = \frac{(1+\varepsilon)K^0+(1-\varepsilon)\bar{K}^0}{\sqrt{2(1+|\varepsilon|^2)}},~~~  i K_S = \frac{(1+\varepsilon)K^0-(1-\varepsilon)\bar{K}^0}{\sqrt{2(1+|\varepsilon|^2)}}\,.
  \label{eq:epsilon}
\end{align}
with\footnote{The mixing parameter appearing in \Eq{eq:epsilon}, sometimes denoted as $\tilde{\varepsilon}$, is, in principle, convention-dependent~\cite{ParticleDataGroup:2024cfk}. For convenience, we work in the Wu-Yang phase convention~\cite{Wu:1964qx} where $\tilde{\varepsilon} = \varepsilon$. }
\begin{align}
    \varepsilon  = 2.228(11)\times 10^{-3}e^{i\theta_{\varepsilon}}\,, \;\;\;\; \theta_{\varepsilon} \approx 0.76\,.
\end{align}
%%%%%%%%%%%%%%%%%%%%%%%%%%%%%%%%%%%%%%
%%%%%%%%%%%%%%%%%%%%%%%%%%%%%%%%%%%%%%
\subsection{Parity and CP}
%%%%%%%%%%%%%%%%%%%%%%%%%%%%%%%%%%%%%%
%%%%%%%%%%%%%%%%%%%%%%%%%%%%%%%%%%%%%%
The structure of the kaon decay amplitudes can be understood using a spurion analysis of the two $Z_2$ symmetries, $P$ and $CP$, defined above the QCD scale as\footnote{Note that for $P$ we omitted a factor of $(-1)^\mu$, where $(-1)^\mu=0$ for $\mu=0$ and is otherwise $1$. This factor is canceled out once the current is coupled to another Lorentz vector with a similar transformation. }
\begin{align}
   P \;\;\;:\;\;\; \bar{q}^i_L  \gamma_\mu q^j_L
     &\leftrightarrow
    \bar{q}^i_R  \gamma_\mu q^j_R\,,
    \\   CP\;\;\; :\;\;\;    \bar{q}^i_{L}  \gamma_\mu q^j_{L}
    &\leftrightarrow
    \bar{q}^j_{L}  \gamma_\mu q^i_{L}\,, \;\;\;   \bar{q}^i_{R}  \gamma_\mu q^j_{R}
    \leftrightarrow
    \bar{q}^j_{R}  \gamma_\mu q^i_{R}  \,.
\end{align}
By identifying $\Sigma\sim \langle \bar{q}_L q_R\rangle$, we find the corresponding transformations below the QCD scale,
\begin{align}
      P \;\;\;:\;\;\; \Pi
     &\to -\Pi\,,
    \\   CP\;\;\; :\;\;\;    \Pi
    &\to 
    -\Pi^*  \,, 
\end{align}
where $\Pi$ is the matrix of the Nambu-Goldstone bosons, defined in \Eq{eq:PI_def}.
In order to keep track of the explicit $P$ and $CP$ violation of the theory, we can promote the ALP couplings to spurions and take the ALP to be odd under both $P$ and $CP$.
Any coupling in the Lagrangian multiplying a $P$- or $CP$-eigenstate combination of operators inherits the corresponding $P$ or $CP$ parity of that operator.
Thus, vector (axial) ALP couplings are odd (even) under $P$.
The real (imaginary) components of the ALP couplings are even (odd) under $CP$.
The mixing parameter $\varepsilon$, defined in \Eq{eq:epsilon}, is odd.
$CP$-odd couplings are the sources of explicit $CP$ violation, with $CP$ properly restored only when they are set to zero.
The $N_8$ coupling defined in Eq. (\ref{eq:N8}), which originates from the SM weak interactions, is maximally $P$ violating as it involves only left-handed fields. We cannot consistently assign a spurionic charge to restore the $P$ symmetry.
In practice, it can be treated as either odd or even under $P$ for the purpose of our spurion analysis.

All the neutral mesons fields are odd under both $P$ and $CP$ with the exception of $K_S$, the only (approximately) $CP$-even state. 
Thus, we can predict that the $CP$-conserving amplitude $K_L \to \pi^0 \pi^0 a$ is proportional to couplings carrying the spurionic charges $\{+,+\}$ under $P$ and $CP$, i.e. the real component of an axial ALP coupling or $N_8$, or their imaginary parts multiplied by $\varepsilon$.
Following a similar argument, we summarize the relevant spurionic charges of all the neutral kaon decay amplitudes in \Tab{tab:P_CP}, together with the couplings we expect to contribute. 
Finally, charged mesons are $P$-odd and are mapped to their oppositely-charged counterparts under $CP$.
As a consequence, their decay amplitudes do not a have well-defined spurionic $CP$ charge, and two- and three-body decays are mediated by $P$-odd (vector) and $P$-even (axial) couplings, respectively. 

\begin{table}   
\begin{center}
\begin{tabular}{ |c |c| c| }
 \hline
  & $P=-1 \;\;(\kappa_V,N_8)$ & $P=+1 \;\; (\kappa_A,N_8)$ \\
  \hline
$CP=+1$  & $K_S\to \pi^0 a$ & $K_L\to \pi^0 \pi^0 a \;\text{ or }\;\pi^+ \pi^- a$ \\ 

   (Re[$c$],\;$\varepsilon\cdot$Im[$c$])  &  &  \\ 
  \hline
$CP=-1$  & $K_L\to \pi^0 a$ & $K_S\to \pi^0 \pi^0 a\;\text{ or }\;\pi^+ \pi^-a$ \\
  (Im[$c$],\;$\varepsilon\cdot$Re[$c$])   & &  \\
  \hline
\end{tabular}
\end{center}
\caption{Classification of the decay amplitudes in terms of $P$ and $CP$. The columns show which couplings are consistent with $P$, while the rows show which couplings are consistent with $CP$, where $c$ can be either coupling appearing in the first row in of the appropriate column.   }
\label{tab:P_CP}
\end{table}
%%%%%%%%%%%%%%%%%%%%%%%%%%%%%%%%%%%%%%
%%%%%%%%%%%%%%%%%%%%%%%%%%%%%%%%%%%%%%
 \subsection{\texorpdfstring{\boldmath $K \to \pi a$}{K to pi a}}
 \label{sec:two_body}
 %%%%%%%%%%%%%%%%%%%%%%%%%%%%%%%%%%%%%%
 %%%%%%%%%%%%%%%%%%%%%%%%%%%%%%%%%%%%%%
The two-body decay rates of $K_{L,S}\to\pi^0a$ and $K^+\to\pi^+a$ are given by
\begin{align}
\label{eq:amp_KL_2body}
  \Gamma(K \to \pi a) = (2m_{K})^{-1}\int \mathrm{d}\Pi_2 |\mathcal{M}(K \to \pi a)|^2 \,, 
\end{align} 
where $\mathrm{d}\Pi_2$ is the usual 2-particle phase space integral. The amplitude for the $K_L$ decay is given by
\begin{align}  
\label{eq:KL_2_body}
\mathcal{M}(K_L \to \pi^0 a)
=& \frac{(m_K^2-m_\pi^2)}{2f} \bigg(    \text{Im}\,\MKpia-i \varepsilon\text{Re}\,\MKpia\bigg)\,,
\end{align}
where we have defined
\begin{align}\label{eq:KLMTilde}
    \MKpia \equiv \kappa_V+ N_8 \mathcal{C}_1\,.
\end{align}
The function $\mathcal{C}_1$ depends on flavor-preserving (i.e. flavor-diagonal and hence \emph{real}) couplings and meson masses, see \App{app:Ci} for the full expression. At leading order in the limit $m_K \gg m_\pi \gg m_a$, we can approximate
\begin{align}
   \mathcal{C}_1 \approx & 
    \;-2 c_{GG}(\Lambda)\, 
    -2 \big[\cdR(\Lambda)\big]_{11}+\big[\cQL(\Lambda)\big]_{11}
+\big[\cQL(\Lambda)\big]_{22}\,.
\end{align}
Our calculation agrees with previous results~\cite{Bauer:2021mvw}.
The appearance of the vector coupling $\kappa_V$ in \Eq{eq:KLMTilde} is a consequence of the $P$ symmetry (see Table (\ref{tab:P_CP})).
The required violation of the $CP$ symmetry is evident in the structure of the amplitude, which is either proportional to an imaginary coupling or to the $CP$-violating $\varepsilon$ parameter and a real coupling.

Similarly, the amplitude for the two-body decay of $K_S$ is given by
\begin{align}  
\label{eq:KS_2_body}
\mathcal{M}(K_S \to \pi^0 a)
=& -\frac{(m_K^2-m_\pi^2)}{2f} \bigg(    \text{Re}\,\MKpia+i \varepsilon  \text{Im}\,\MKpia 
\bigg)\,,
\end{align} 
Similarly to $K_L$, the structure of this amplitude is also compatible with $P$ and $CP$ symmetries as shown in  \Tab{tab:P_CP}.
Finally, the charged decay amplitude is given by
\begin{align}
    \mathcal{M}(K^+ \to \pi^+ a) =& \frac{i(m_K^2-m_\pi^2)}{2f} \bigg(    \MKpia+N_8\Delta \mathcal{C}_1\bigg)\,,
\end{align} 
where
\begin{align}
        \Delta \mathcal{C}_1 = \frac{m_\pi^2 (c_d-c_u)}{(m^2_\pi-m_a^2)  }\,,~~{\rm{with}}~~c_{q} \equiv k_{q}-k_{Q}.
\end{align}
%%%%%%%%%%%%%%%%%%%%%%%%%
%%%%%%%%%%%%%%%%%%%%%%%%%
\subsection{\texorpdfstring{\boldmath $K \to \pi^0 \pi^0 a$}{K to pi0 pi0 a}}
\label{Sec:Kpi0pi0}
%%%%%%%%%%%%%%%%%%%%%%%%%
%%%%%%%%%%%%%%%%%%%%%%%%%
The three-body decay rate of $K_L$ and $K_S$ to the fully neutral final state is given by
\begin{align}
\label{eq:amp_KL_3body}
  \Gamma(K\to \pi^0 \pi^0 a) = (2m_{K})^{-1}\int \frac12\mathrm{d}\Pi_3 |\mathcal{M}(K\to \pi^0 \pi^0 a)|^2  \,,
\end{align} 
where $\mathrm{d}\Pi_3$ is the usual 3-particle phase space integral and the factor $1/2$ accounts for the identical particles in the final state.
The amplitude for $K_L$ is
\begin{align}  
\label{eq:KL_3_body_neutral}
    &\mathcal{M}(K_L \to a\, \pi^0 \,\pi^0)  
=  \frac{m_K^2}{2 \sqrt{2} f_{\pi} f}\left(\text{Re}\, \mathcal{M}_0+i\varepsilon \text{Im}\, \mathcal{M}_0 \right)\,,
\end{align}
where we have defined
\begin{align}
\label{eq::M0_def}
    \mathcal{M}_0 \equiv  \kappa_A\left(\frac{s_{\pi^0}}{m_K^2-m_{a}^2} -1\right)+N_8
        \left( \mathcal{C}_2+\mathcal{C}_3\frac{s_{\pi^0}}{m_K^2}\right)\,.
\end{align}
The kinematical variable $s_{\pi^0}$  is defined as $s_{\pi^0} \equiv (p_{\pi^0_1}+p_{\pi^0_2})^2$, with $p_{\pi^0_1}$ and $p_{\pi^0_2}$ the 4-momenta of the pions in the final state.
At this order in the momentum expansion, the $\pi^0$ exchange symmetry ensures the amplitude depends only on this kinematic variable, while higher order terms in $\chi$PT may produce dependence on additional independent kinematic variables.
The appearance of the axial combination $\kappa_A$ in \Eq{eq::M0_def} can be understood as a consequence of the $P$ symmetry (see Table (\ref{tab:P_CP})).
$CP$ symmetry is evident in the structure of the amplitude, which is either proportional to a real coupling or to a product of an imaginary coupling and the $CP$-violating $\varepsilon$ parameter.

The coefficients $\mathcal{C}_2$ and $\mathcal{C}_3$ which multiply the weak parameter $N_8$ are functions of flavor-preserving (i.e. flavor-diagonal and hence \emph{real}) couplings and meson masses, see \App{app:Ci} for the full expressions.
This contribution has not been discussed recently when considering three-body decays~\cite{MartinCamalich:2020dfe,Cavan-Piton:2024pqp}.
At leading order in the limit $m_K \gg m_\pi\gg m_a$, we can approximate
\begin{align}
   \mathcal{C}_2 \approx & 
    \;-2 c_{GG}(\Lambda)\, 
    -2 \big[\cdR(\Lambda)\big]_{11}+\big[\cQL(\Lambda)\big]_{11}
+\big[\cQL(\Lambda)\big]_{22}
\,,
    \\
    \mathcal{C}_3 \approx & 
    -3\big[\cQL(\Lambda)\big]_{11}
     +2\big[\cdR(\Lambda)\big]_{22}
    +\big[\cQL(\Lambda)\big]_{22}\,.
\end{align}

In light of \Eq{eq:KL_3_body_neutral}, the structure of the amplitude for the three-body $K_S$ decay is not surprising.
It is given by
\begin{align}  
\label{eq:KS_3_body_neutral}
    &\mathcal{M}(K_S \to a\, \pi^0 \,\pi^0)  
=  \frac{m_K^2}{2 \sqrt{2} f_{\pi} f}\left(\text{Im}\, \mathcal{M}_0-i\varepsilon \text{Re}\, \mathcal{M}_0 \right)\,.
\end{align}
Before concluding this section, we observe that, unlike for the two-body decay calculation, our three-body decay calculation requires the inclusion of naively factorizable\footnote{In this context, factorizable means amplitudes which can be factorized into products of lower-point amplitudes  with at least 3 external states across an internal propagator.} contributions, 
which cancel out non-physical basis-dependent contributions originating from contact terms. 
In fact, we note that by summing up only the contact terms, one finds 
\begin{align}
&\mathcal{M}_{\text{\tiny contact}}(K_L \to a\, \pi^0 \,\pi^0)
\nonumber
\\\label{eq:nonphys3b}
&= \frac{\sqrt{2} c_{GG} \text{Re}\,N_8(m_{\pi}^2 -s_{\pi^0})}{f_\pi f}(\delta_s-\delta_d)+(\text{basis-independent terms})\,,
\end{align}
which is not physical as clearly shown by the dependence on the basis-dependent parameters $\delta_s$ and $\delta_d$ defined in Eq. (\ref{eq:generic_basis}).
%%%%%%%%%%%%%%%%%%%%%%
\begin{figure}[t]
  \centering
    \includegraphics[width=0.9\textwidth]{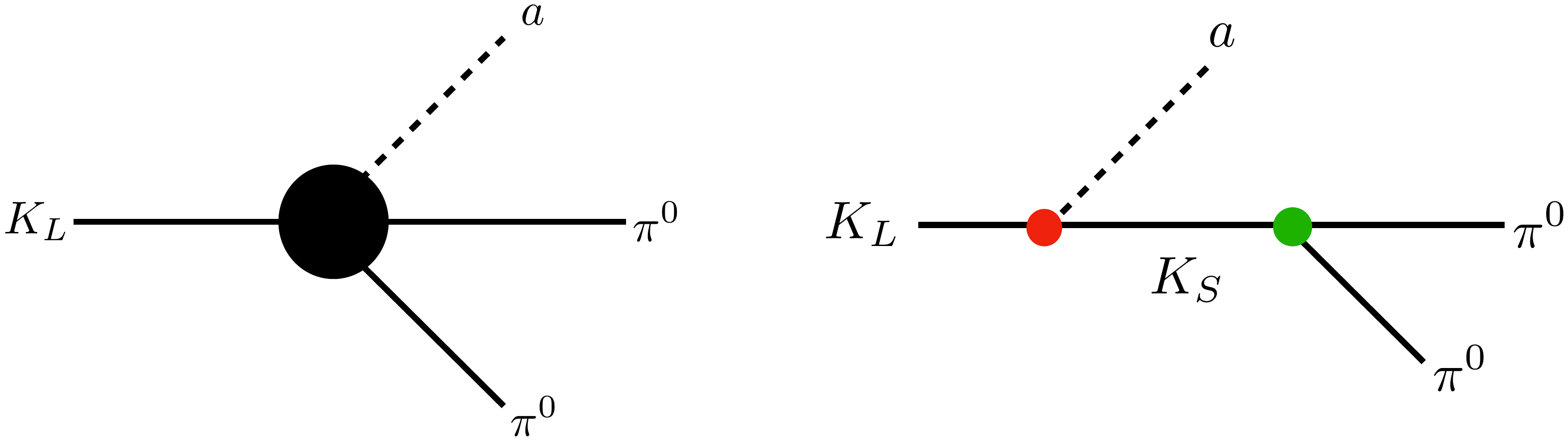}
      \caption{Contributing diagrams to $K_L \to \pi^0 \pi^0 a$. 
      The non-physical contribution from contact term (left) is canceled by diagram mediated by the $K_S$ (right), see text for more details.}
      \label{fig:diagrams}
\end{figure}
%%%%%%%%%%%%%%%%%%%%%%
This issue is resolved by an additional contribution, coming from a diagram in which the decay process is mediated by $K_S$, see \Fig{fig:diagrams}.
It involves two 3-point interactions; one from the kinetic term,
\begin{align}
\label{eq:non_phys_int}
&\mathcal{L}_{\text{\tiny kin}}  \supset 
\frac{c_{GG}}{ f} (  \delta_d-\delta_s) (\partial_\mu a)(K_S\partial^\mu K_L - K_L \partial^\mu K_S)\,,
\end{align}
illustrated by the red vertex in \Fig{fig:diagrams},
and another from the weak-induced interactions,
\begin{align}   &\mathcal{L}_{\text{\tiny weak}}  \supset\frac{\text{Re}\,N_8}{\sqrt{2}f_\pi}(\partial_\mu \pi^0) (K_S \partial^\mu \pi^0- \pi^0 \partial^\mu K_S )\,.
\end{align}
illustrated by the green vertex in \Fig{fig:diagrams}.
We note that the interaction vertex in \Eq{eq:non_phys_int}
contains the same non-physical phase combination as in Eq. (\ref{eq:nonphys3b}).
Indeed, it does not contribute to any physical process where all three particles are on-shell. In fact, in the on-shell case it is proportional to $m^2_{K_L}-m^2_{K_S}$, which vanishes at leading order in $\chi$PT.
Carefully calculating the naively factorizable contribution, one finds that it is in fact secretly a contact term as well: the interaction vertex exactly cancels out the $K_S$ propagator, resulting in
\begin{align}
&\mathcal{M}_{\text{\tiny factor.}}(K_L \to a\, \pi^0 \,\pi^0) \nonumber
\\
&= 
-\frac{\sqrt{2} c_{GG} |N_8|\cos \delta_8(m_{\pi}^2 -s_{\pi^0})}{f_\pi f}(\delta_s-\delta_d)+(\text{basis-independent terms})\,.
\end{align}
The basis-dependent contributions cancel out, leading to our basis-independent final result.

%%%%%%%%%%%%%%%%%%%%%%%%%%%%%
%%%%%%%%%%%%%%%%%%%%%%%%%%%%%
\subsection{\texorpdfstring{\boldmath $K \to \pi^+ \pi^- a$}{K to pi+ pi- a}}
\label{Sec:Kpi+pi-}
%%%%%%%%%%%%%%%%%%%%%%%%%%%%%
%%%%%%%%%%%%%%%%%%%%%%%%%%%%%
The three-body decay rates of $K_L$ and $K_S$ to the charged final state are given by
\begin{align}
\label{eq:amp_KL_3body_Charged}
  \Gamma(K \to \pi^+ \pi^- a) = (2m_{K})^{-1}\int\mathrm{d}\Pi_3 |\mathcal{M}(K \to \pi^+ \pi^- a)|^2  \,.
\end{align} 
Since the $\pi^+ \pi^- a$ final state does not contain identical particles, the amplitude depends on two kinematic variable, which we take to be
\begin{align}\label{Eq:kinvariables}
    s_{\pi} \equiv (p_{\pi^+}+p_{\pi^-})^2\,, \;\; \tilde{s} = p_K \cdot (p_{\pi^-}-p_{\pi^+}) = p_a \cdot (p_{\pi^-}-p_{\pi^+})\,,
\end{align}
where we can also write $\tilde{s} = 2 (p_{\pi^+}+p_a)^2+s_{\pi}-m_{a}^2-m_K^2-2 m_\pi^2$.
The variable $\tilde{s}$ is a particularly suitable choice for a kinematic variable,  with well-defined transformation properties under $CP$, which exchanges $\pi^+ \leftrightarrow \pi^-$ and therefore $\tilde{s}\to -\tilde{s}$. 
The other Lorentz invariants, i.e. $s_\pi$ and the masses, are even under $CP$. 

The $K_L$ amplitude is then given by
\begin{align}
\label{eq:KL_3_body_charged}
    \mathcal{M}(K_L \to a\, \pi^+ \,\pi^-)
        =\frac{m_K^2}{2 \sqrt{2} f_\pi f}\bigg(\text{Re}\,\mathcal{M}_+
        -i\text{Im}\,\mathcal{M}_-+i\varepsilon[ \text{Im}\,\mathcal{M}_+ +i  \text{Re}\,\mathcal{M}_- ]\bigg)\,,
\end{align}
with
\begin{align}
    \mathcal{M}_{+} &\equiv  \kappa_A\left(\frac{s_{\pi}}{m_K^2-m_{a}^2} -1\right)+  N_8\left( \mathcal{C}_4+\mathcal{C}_5\frac{s_\pi}{m_K^2}\right)\,,
    \\
    \label{eq:M-S}
        \mathcal{M}_{-} &\equiv   \left(\frac{
 \kappa_A }{ m_K^2-m_a^2}+   \frac{ N_8\mathcal{C}_6}{m_K^2}\right) \tilde{s}\,.
\end{align}
$\mathcal{C}_4,\mathcal{C}_5$ and $\mathcal{C}_6$ are functions of flavor-preserving (i.e. flavor-diagonal and hence \emph{real}) couplings and meson masses, which we calculated for the first time. See \App{app:Ci} for the full expressions.
At leading order in the limit $m_K \gg m_\pi \gg m_a$, we can approximate
\begin{align}
\mathcal{C}_4\approx & 
-2 c_{GG}-\big[\cdR(\Lambda)\big]_{11}+\big[\cQL(\Lambda)\big]_{11}+\big[\cQL(\Lambda)\big]_{22}-\big[\cuR(\Lambda)\big]_{11}
\\
 \mathcal{C}_5\approx & -\big[\cdR(\Lambda)\big]_{11}+2 \big[\cdR(\Lambda)\big]_{22}-3 \big[\cQL(\Lambda)\big]_{11}+\big[\cQL(\Lambda)\big]_{22}+\big[\cuR(\Lambda)\big]_{11}
\\
\mathcal{C}_6\approx &-2 c_{GG} -\big[\cdR(\Lambda)\big]_{11}+\big[\cQL(\Lambda)\big]_{11}+\big[\cQL(\Lambda)\big]_{22}-\big[\cuR(\Lambda)\big]_{11}\,.
\end{align}
As for the neutral final state, this three-body decay calculation also requires the inclusion of naively factorizable terms, which cancel out non-physical contributions originating from contact terms.
In this case, in addition to the diagram in which $K_S$ mediates the process, we must also include two additional diagrams in which the process is mediated by $\pi^+$ and $\pi^-$.

The amplitude for $K_S$ decays in given by
\begin{align}
\label{eq:KS_3_body_charged}
    \mathcal{M}(K_S \to a\, \pi^+ \,\pi^-)
    =\frac{m_K^2}{2 \sqrt{2} f_\pi f}\bigg(\text{Im}\,\mathcal{M}_+ +i\text{Re}\,\mathcal{M}_- -i\varepsilon[ \text{Re}\,\mathcal{M}_+ -i \text{Im}\,\mathcal{M}_- ]\bigg)\,.
\end{align}
Interestingly, in the SM the three-body decay $K_S \to \pi^+ \pi^- \pi^0$ is not mediated by the octet operator considered in this work, but by the ${\bf 27}$ operator~\cite{Bijnens:2002vr}.
This can be understood as a consequence of an  isospin selection rule: the octet operator carries $\Delta I{=}1/2$ isospin charge. Therefore, the initial configuration with the $I{=}1/2$ kaon can only decay to either a $I{=}0$ or a $I{=}1$ final state.
At the same time, the only non-trivial representation for the final state involving three pions is $I{=}1$,
 since the $I{=}0$ representation vanishes due to Bose symmetry.
 The $I{=}1$ representation is symmetric under pion exchange.
The $CP$ symmetry exchanges the charged pion states. For this reason, in the absence of CP violation the octet contributes to the $K_L\to\pi^+\pi^-\pi^0$ but not to the $K_S\to\pi^+\pi^-\pi^0$ decay.

This argument was used to claim that the octet operator does not contribute to $K_S \to \pi^+ \pi^- a$~\cite{Cavan-Piton:2024pqp}.
However, since the ALP is not part of an isospin representation, this argument does not apply when replacing $\pi^0$ with an ALP. 
Treating the ALP as an isospin singlet, the two-pion system can be in an $I{=}1$ configuration which contains the anti-symmetric combination expected for the $K_S$ decay in the absence of $CP$ violation.
Indeed, without $CP$ violation one finds $\mathcal{M}(K_S \to a\, \pi^+ \,\pi^-) \propto \text{Re}\, \,\mathcal{M}_-$, i.e. the amplitude receives a contribution from the octet operator (see Eq. (\ref{eq:M-S}) for the expression for $\mathcal{M}_-$).

%%%%%%%%%%%%%%%%%%%%%%%%%%%%%
%%%%%%%%%%%%%%%%%%%%%%%%%%%%%
\section{Kaon rates}
\label{sec:rates}
%%%%%%%%%%%%%%%%%%%%%%%%%%%%%
%%%%%%%%%%%%%%%%%%%%%%%%%%%%%
\subsection{Flavor-changing couplings}
\label{sec:FV_couplings}
\begin{figure}[t]
  \centering
    \includegraphics[width=0.9\textwidth]{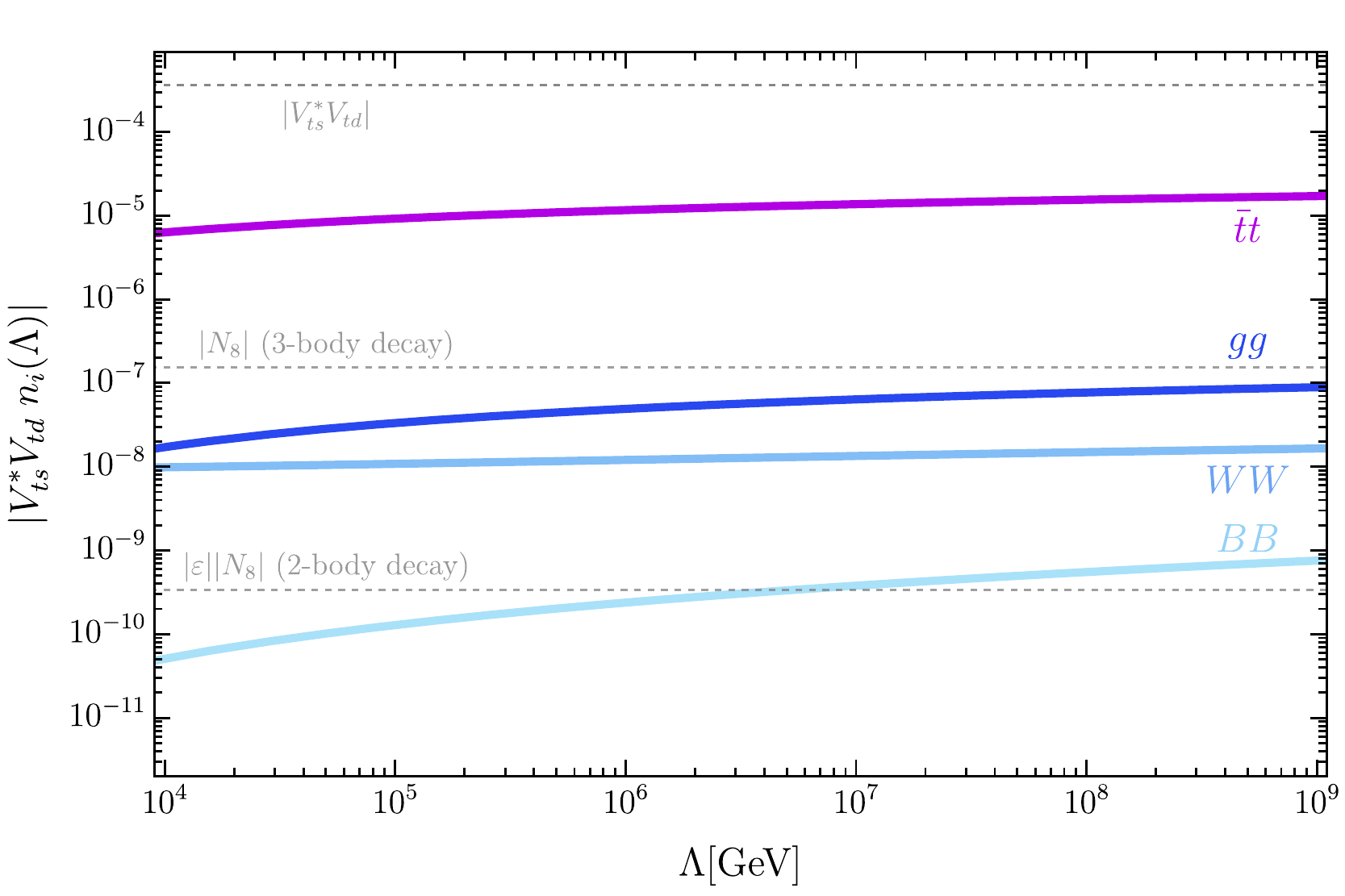}
      \caption{The numerical coefficients of $c_{tt}(\Lambda),\tilde{c}_{GG}(\Lambda),\tilde{c}_{WW}(\Lambda),\tilde{c}_{BB}(\Lambda)$ as a function of the UV scale $\Lambda$ (see Eqs. (\ref{eq:KL}), (\ref{eq:rad_correction})). We also show for reference the values of $|V^*_{td}V_{ts}|$, $|N_8|$ and $|\varepsilon||N_8| $.
      The first dominates the flavor-violating coupling for $c^{\text{\tiny NP}}_L \sim 1$, while the last two are the coefficients of the flavor-diagonal couplings contributing for example to the $K_L$ three-body and two-body decays involving neutral pions, respectively. }
      \label{fig:couplings}
\end{figure}

The low energy flavor-changing couplings play a crucial role in any discussion about kaon decay rates.
The flavor-changing couplings can be written as,
\begin{align}
    \kappa_R &= \kappa^{\text{\tiny FV}}_R+\kappa^{\text{\tiny MFV}}_R \,,
    \\
    \kappa_L &= \kappa^{\text{\tiny FV}}_L+\kappa^{\text{\tiny MFV}}_L\,.
\end{align}
The first terms are present in UV theories which contain new sources of flavor violation, namely theories which violate the MFV hypothesis. 
The second terms are generated in MFV theories, either by SM radiative corrections, see Eq. (\ref{eq:KL}), or by NP contributions.
Under the MFV hypothesis, the ALP--quark coupling matrices furnish the following representations of the $SU(3)_{Q_L}\times SU(3)_{u_R} \times SU(3)_{d_R}$ flavor group,
\begin{align}
    {\bf c}_{Q_L} \sim ({\bf 8,1,1})\,,
    \\
    {\bf c}_{u_R} \sim ({\bf 1,8,1})\,,
    \\
    {\bf c}_{d_R} \sim ({\bf 1,1,8})\,.
\end{align}
We construct the MFV expansion using the SM Yukawa matrices, which transform (spurionically) as,
\begin{align}
    {\bf Y}_u \sim ({\bf 3,\bar{3},1})\,,
    \\
    {\bf Y}_d \sim ({\bf 3,1,\bar{3}})\,.
\end{align} 
One finds the following expansion in Yukawa matrices,
\begin{align}\label{eq:cQMFV}
    {\bf c}_{Q} &= c_0^{Q} {\bf 1} + \left[c_{1,1}^{Q}{\bf Y}_u{\bf Y}^\dagger_u+c_{1,2}^{Q}{\bf Y}_d{\bf Y}^\dagger_d\right]+ \left[c_{2}^{Q}{\bf Y}_u{\bf Y}^\dagger_u{\bf Y}_d{\bf Y}^\dagger_d+\text{h.c}\right]+\cdots\,,
    \\
    {\bf c}_{u_R} &= c_0^{u} {\bf 1} + c_{1}^{u}{\bf Y}^\dagger_u{\bf Y}_u+ \left[c_{2,1}^{u}{\bf Y}^\dagger_u{\bf Y}_u{\bf Y}^\dagger_u{\bf Y}_u+c_{2,2}^{u}{\bf Y}^\dagger_u{\bf Y}_d{\bf Y}^\dagger_d{\bf Y}_u\right]+\cdots\,,
      \\
    {\bf c}_{d_R} &= c_0^{d} {\bf 1} +  c_{1}^{d}{\bf Y}^\dagger_d{\bf Y}_d+ \left[c_{2,1}^{d}{\bf Y}^\dagger_d{\bf Y}_d{\bf Y}^\dagger_d{\bf Y}_d+c_{2,2}^{d}{\bf Y}^\dagger_d{\bf Y}_u{\bf Y}^\dagger_u{\bf Y}_d\right]+\cdots\,,
\end{align}
where ${\bf 1}$ is the $3\times 3$ identity matrix and the $c$ are arbitrary flavor-blind parameters.
We note that the first two sets of terms in the MFV expansion do not introduce additional phases due to the hermiticity of the coupling matrices ${\bf c}_{Q}$, ${\bf c}_{u_R}$, and ${\bf c}_{d_R}$ and, therefore, there are no new sources of CPV. Terms further suppressed by higher powers of Yukawa couplings can introduce new sources of CPV. For example,
the coefficient $c_{2}^Q$ appearing in the expansion of ${\bf c}_{Q}$ can in principle be complex. Also the expansions for ${\bf c}_{u_R}$ and ${\bf c}_{d_R}$ can contain new sources of CPV. An example is the term ${\bf Y}^\dagger_u{\bf Y}_u{\bf Y}^\dagger_u{\bf Y}_d{\bf Y}^\dagger_d{\bf Y}_u$ in the expansion of ${\bf c}_{u_R}$.
In order to find the structure of the leading MFV contributions, we consider the leading order terms in the spurion expansion which contains off-diagonal terms, namely
\begin{align}
     [{\bm c}_{d_R}]_{12} &\supset 
   [{\bm Y}^\dagger_d 
    {\bm Y}_u 
    {\bm Y}^\dagger_u 
    {\bm Y}_d]_{12}  = [\hat{{\bm Y}}^\dagger_d {\bm V}^\dagger
    \hat{{\bm Y}}_u 
    \hat{{\bm Y}}^\dagger_u 
    {\bm V}\hat{{\bm Y}}_d]_{12} =   V_{ts}V^*_{td} y_t^2 y_s y_d+...\,,
    \\
           \big[{\bm V}^\dagger \cQL{\bm V}\big]_{12}
           &\supset  \big[{\bm V}^\dagger {\bm Y}_u{\bm Y}^\dagger_u{\bm V}\big]_{12}
           = {V}^*_{td}{V}_{ts}y_t^2 +...\,,
\end{align}
where we omit terms which are subleading due to CKM or Yukawa suppression.
Here we used the fact that in our convention ${\bm Y}_u = \hat{\bm Y}_u \equiv \text{Diag}(y_u,y_c,y_t)$ and ${\bm Y}_d = {\bm V}\hat{\bm Y}_d \equiv {\bm V}\text{Diag}(y_d,y_s,y_b)$.
Thus, in the MFV case, instead of the generic couplings in Eq. (\ref{eq:KL}), we can write,
\begin{align}
    \kappa^{\text{\tiny MFV}}_R &\equiv  V_{ts}V_{td}^*y_s y_d \,c_{R}^{\text{\tiny MFV}} \equiv V_{ts}V_{td}^*y_s y_d( c_{R}^{\text{\tiny NP}}+c_{R}^{\text{\tiny SM}})\,,
    \\
    \kappa^{\text{\tiny MFV}}_L &\equiv V_{ts}V^*_{td}\,c_{L}^{\text{\tiny MFV}}\equiv  V_{ts}V^*_{td}(c_{L}^{\text{\tiny NP}}+c_{L}^{\text{\tiny SM}})\,,
\end{align}
where we took $y_t \approx 1$.
Importantly, we take the coefficients $\{c_{R}^{\text{\tiny NP}},c_{R}^{\text{\tiny SM}},c_{L}^{\text{\tiny NP}},c_{L}^{\text{\tiny SM}}\}$ to be real, as expected by the leading order MFV expansion.
Along with $N_8$ and $\varepsilon$, the coupling combination ${V}^*_{td}{V}_{ts}$ is the last source of flavor and $CP$-violation relevant to our discussion. 
Numerically,
\begin{align}
|{V}^*_{td}{V}_{ts}| =  3.6 \times 10^{-4} \,, \;\;\; \text{Arg}[{V}^*_{td}{V}_{ts}] = \frac{\eta}{\rho-1}=-0.39\,,
\end{align}
where $\eta=0.348$ and $\rho=0.159$ are the  conventional Wolfenstein parameters~\cite{ParticleDataGroup:2024cfk}.
We find that $\kappa_R^{\text{\tiny MFV}}/\kappa_L^{\text{\tiny MFV}} = y_s y_d(c^{\text{\tiny MFV}}_R/c^{\text{\tiny MFV}}_L)$ and we can safely neglect the right-handed MFV coupling as long as $c^{\text{\tiny MFV}}_R \ll c^{\text{\tiny MFV}}_L/(y_s y_d)$.
In that case, we can write
\begin{align}
\label{eq:Kappa_V}
    \kappa_{V} &\approx \kappa^{\text{\tiny FV}}_{V}+ {V}^*_{td}{V}_{ts}( c^{\text{\tiny NP}}_L+c_L^{\text{\tiny SM}})\,,
    \\
    \label{eq:Kappa_A}
        \kappa_{A} &\approx \kappa^{\text{\tiny FV}}_{A}- {V}^*_{td}{V}_{ts}( c^{\text{\tiny NP}}_L+c_L^{\text{\tiny SM}})\,,
\end{align}
where we identify 
\begin{align}
    c_L^{\text{\tiny NP}} = \big[  \cQL(\Lambda)  \big]_{33}-\big[  \cQL(\Lambda)  \big]_{11}\,,\;\;\;\; c_L^{\text{\tiny SM}} = \Delta \kappa_L\,,
\end{align}
see \Eq{eq:UV_off_diag_couplings} and \Eq{eq:rad_correction}.
In \Fig{fig:couplings} we plot the coefficients of the UV couplings appearing in $\Delta \kappa_L$ as a function of $\Lambda$.
We plot for reference $|V^*_{ts}V_{td}|$, $N_8$ and $\varepsilon\,N_8$ as gray horizontal lines.
The last two couplings contribute to the $K_L$ three-body and two-body decay involving neutral pions, respectively.

%%%%%%%%%%%%%%%%%%%%%%

We define the ratios
\begin{align}
\label{eq:ratio_defs}
    R_{0} \equiv \frac{\text{Br}(K_L \to \pi^0 \pi^0 a)}{\text{Br}(K_L \to \pi^0 a)}
\;\;\;\text{  and  }\;\;\;
    R_{\pm}& \equiv \frac{\text{Br}(K_L \to \pi^+ \pi^- a)}{\text{Br}(K_L \to \pi^0 a)}\,.
\end{align}
Schematically,
\begin{align}
    R_0,R_\pm
    \sim\left|
    \frac{\text{Largest FV+$CP$-preserving coupling}}{\text{Largest FV+$CP$-violating coupling}}
    \right|^2 \cdot I_{0,\pm}(m_a)\,,
    \label{eq:ratio_schematic}
\end{align}
where $I_{0,\pm}(m_a)$ denote the phase space ratios,
\begin{align}\label{Eq:I}
    I_{0,\pm}(m_a) \sim \frac{\text{three-body phase space}}{\text{two-body phase space}}\sim \mathcal{O}(10^{-3}-10^{-2})\,,
\end{align}
that are defined explicitly in App. \ref{app:PSI_and_ratios}. The numerator in \Eq{eq:ratio_schematic} is sensitive to the largest $CP$-preserving source of flavor violation, while the denominator is sensitive to the largest $CP$-violating source of flavor violation.
These sources may have different origins, which could lead to different scaling.
Let us classify all the different possible hierarchies.
To facilitate this classification, we consider the coupling parameterization of this section.
Since $P$ is severely violated by all the flavor-changing couplings, we can safely assume that any new source of FV would also violate $P$ explicit such that $\kappa^{\text{\tiny FV}}_V \approx \kappa^{\text{\tiny FV}}_A \equiv \kappa^{\text{\tiny FV}} $ (see \Eq{eq:Kappa_V} and \Eq{eq:Kappa_A}),
and omit the $V/A$ subscripts for the remainder of the discussion.
\subsection{Maximal flavor violation}
\label{sec:max_FV}
The first class of theories we consider are schematically defined by,
\begin{align}
\kappa^{\text{\tiny FV}}\gg \text{Max}\big[\kappa_L^{\text{\tiny MFV}},N_8\big]\,,
\end{align}
which we dub as \emph{maximal flavor violation}.
It should be understood that $N_8$ is always accompanied by flavor-diagonal couplings. We omit them here and in the following discussion for brevity.
Both the two- and three-body decays are dominated by the same coupling $\kappa^{\text{\tiny FV}}$, and the ratio of rates depends on the alignment of this coupling with CP-violation, namely on the phase $\theta_{\kappa} \equiv \text{Arg}\,\kappa^{\text{\tiny FV}}$.
For the neutral ratio, we find the scaling
\begin{align}\label{eq:maximallyR0}
   R_0 \sim I_0(m_a) \cdot \begin{cases}
       1 \;\;\;\;\;\; & \tan \theta_{\kappa}\sim \mathcal{O}(1)
       \\
       \left|\frac{1}{\varepsilon}\right|^2 & \tan \theta_{\kappa}\ll |\varepsilon| 
              \\
    \left|\varepsilon\right|^2 & \cot \theta_{\kappa}\ll |\varepsilon| 
   \end{cases}\,.
\end{align}
The ratio is schematically determined by the phase space ratio $I_0(m_a)$ for a coupling with moderate $CP$ violation $\theta_{\kappa}\sim \mathcal{O}(1)$, is enhanced by $1/|\varepsilon|^2$ if the coupling is approximately $CP$-preserving and is suppressed by $|\varepsilon|^2$ if the coupling is approximately imaginary. 
For the charged ratio,
\begin{align}\label{eq:maximallyRpm}
   R_\pm \sim I_\pm(m_a) \cdot \begin{cases}
       1 \;\;\;\;\;\; & \tan \theta_{\kappa}\sim \mathcal{O}(1)
       \\
       \left|\frac{1}{\varepsilon}\right|^2 & \tan \theta_{\kappa}\ll |\varepsilon| 
              \\
    1 & \cot \theta_{\kappa}\ll |\varepsilon| 
   \end{cases}\,.
\end{align}
The charged ratio shows the same scaling as the neutral ratio except for the case of an approximately imaginary coupling. In this case, the three-body amplitude contains a contribution  proportional to $\text{Im}\kappa^{\text{FV}} \tilde{s}$, which is absent from the neutral three-body decay.
As a consequence, the charged ratio is determined by the phase space ratio $I_\pm(m_a)$ also for an approximately imaginary coupling. 
More details on the full numerical calculation of the ratios are provided in \App{sec:non_MFV}.

\begin{figure}[t]
  \centering    \includegraphics[width=0.8\textwidth]{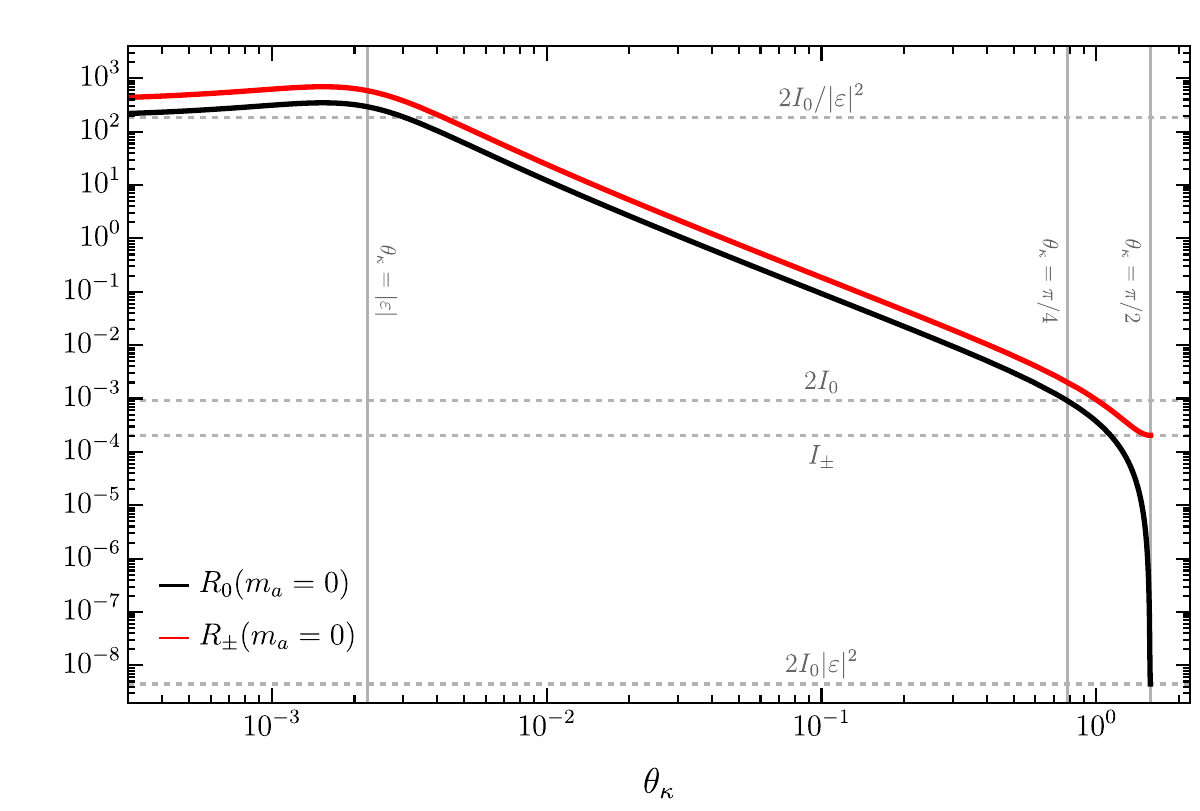}
      \caption{$R_0$ and $R_\pm$ as a function of $\text{Arg}\,\kappa^{\text{\tiny FV}}\equiv \theta_{\kappa}$ plotted in black and red, respectively. 
      The dashed horizontal lines mark the approximations in the various cases (see Eqs. (\ref{eq:maximallyR0}), (\ref{eq:maximallyRpm})). 
      Except around $\theta_\kappa \approx \pi/2$, the ratios differ by a factor of two due to the $\pi^0$ exchange symmetry, see \Eq{eq:amp_KL_3body} and \Eq{eq:amp_KL_3body_Charged}.}
      \label{fig:Non_MFV_ratios}
\end{figure}
We summarize our results for the maximal flavor violation theories in \Fig{fig:Non_MFV_ratios}, where we plot $R_0$ ($R_\pm$) in black (red) as a function of $\theta_{\kappa}$, in the case of a very light ALP ($m_a\sim 0$).
We find good agreement with the approximations in the various cases plotted as horizontal dashed gray lines. 
We find that the ratio of ratios $R_{\pm}/R_{0}\approx 2$, except for the region
where the coupling is predominantly imaginary, $\theta_k\sim \pi/2$ (see red vs. black curves in the figure).

\subsection{Minimal flavor violation}
\label{sec::MFV}
The second class of theories we consider are MFV theories, schematically defined by
\begin{align}
\text{Min}\big[\kappa_L^{\text{\tiny MFV}},N_8\big]\gg \kappa^{\text{\tiny FV}} \,.
\end{align}
In these theories, the dominant source of flavor violation is the SM Yukawas, in accordance with the MFV hypothesis.
The ratio of rates is then given schematically by,
\begin{align}
    R_0,R_{\pm} \sim \left|\frac{\text{Max}[\text{Re}\,\kappa_L^{\text{\tiny MFV}},N_8]}{\text{Max}[\text{Im}\,\kappa_L^{\text{\tiny MFV}},\varepsilon\,N_8]}\right|^2\cdot I_{0,\pm}(m_a).
\end{align}
The two- and three-body decays are mediated in most cases\footnote{The only  exception is if $\kappa_L^{\text{MFV}}$ is dominated by $c_L^{\text{NP}}$, which could be generated in principle by another type of interaction.} by the weak interactions.
We say that a process is mediated by the weak interaction \emph{directly} if $\kappa_L^{\text{\tiny MFV}}$ is the dominant coupling, or  \emph{indirectly} if $N_8$ is the dominant coupling, where it is understood that $N_8$ is always accompanied by flavor-diagonal couplings appearing in $\mathcal C_i$. 
For concreteness, let us assume for the remainder of this section that all the fermions couplings are flavor-blind i.e. ${\bm c}_F = c_{F}1_{3\times 3}$ for $F\in\{Q_L,u_R,d_R,L_L,e_R\}$.

1st scenario: if both rates are mediated \emph{directly} by the weak interaction, i.e.  $\kappa_L^{\text{MFV}}\gg N_8\mathcal{C}_i$,  we find the sharp prediction,
\begin{align}\label{eq:RMFV1}
    R_0,R_{\pm} \sim \left(\frac{\text{Re}\, V^*_{td}}{\text{Im}\,V^*_{td}} \right)^2\,I_{0,\pm}(m_a) \sim 10\, I_{0,\pm}(m_a) \,.
\end{align}
Thus, in this case the ratio of rates is only one order of magnitude larger than the naive prediction based on the phase space ratios. 
This scenario can be realized in two ways, (1) by having $c_{tt}(\Lambda)\neq0$ (see \Fig{fig:couplings}) or (2) by completely decoupling the ALP from the strong sector $c_{GG}(\Lambda)=c_{u_R}(\Lambda)= c_{d_R}(\Lambda)=c_{Q_L}(\Lambda)=0$.
In the latter case, we have $\mathcal{C}_i=0$ and the indirectly-mediated process is suppressed. 

2nd scenario: for smaller values of $\kappa_L^{\text{MFV}}$, the three-body rate could be mediated \emph{indirectly} while the two-body rate is still mediated \emph{directly}, namely when $ \varepsilon\, N_8\mathcal{C}_i\ll \kappa_L^{\text{MFV}} \ll N_8 \mathcal{C}_i$.
In this case,
\begin{align}\label{eq:RMFV2}
    R_0,R_{\pm} \sim \left(\frac{N_8}{\text{Im}\,\kappa_L^{\text{MFV}}} \right)^2\,I_{0,\pm}(m_a) \,.
\end{align}
Depending on the magnitude of $\text{Im}\,\kappa_L^{\text{MFV}}$, a large  enhancement of the three-body rate compared to the two-body rate is possible. 
This scenario is realized in theories in which $c_{tt}(\Lambda)=0$ while $\tilde{c}_{GG}(\Lambda)$, $\tilde{c}_{WW}(\Lambda)$ or $\tilde{c}_{BB}(\Lambda)$ is non-vanishing, with the latter depending on the UV scale; see \Fig{fig:couplings}.
A simple and motivated realization of $\tilde{c}_{GG}(\Lambda) \neq0$ is an ALP coupled exclusively to gluons in the UV theory $c_{GG}\neq0$, in which case the coupling hierarchy is
\begin{align}\label{eq:MFVcGG}
        \left(\frac{N_8}{\text{Im}\,n_G V^*_{td}V_{ts}}\right)^2  \approx \begin{cases}
            600 \;\;\;\;\;\;\;\; &\Lambda = 10^4\,\text{GeV}
            \\
            20 & \Lambda = 10^{10}\,\text{GeV}
        \end{cases}\,.
\end{align}
Thus, we find that for low UV scales, the coupling hierarchy is sufficient to overcome the phase-space suppression leading to $R_{0},R_{\pm} \sim 1-10$.
An even larger enhancement is possible if only $\tilde{c}_{WW}(\Lambda)\neq 0$,
 \begin{align}\label{eq:MFVcWW}
     \left(\frac{N_8}{\text{Im}\,n_W V^*_{td}V_{ts}}\right)^2 \approx 1600\,,
 \end{align}
where this result is largely independent on the NP scale, $\Lambda$ (see \Fig{fig:couplings}).
The last realization of this scenario is for high UV scales when $\tilde{c}_{BB}(\Lambda)$ is the only non-vanishing coupling, in which case the largest hierarchy of this scenario is found,
\begin{align}
   \left(\frac{N_8}{\text{Im}\,n_B V^*_{td}V_{ts}}\right)^2  \approx  1.6\times 10^5\,,  
   \label{eq::BB_ratio}
\end{align}
where we evaluated the ratio for $\Lambda = 10^{10}\,\text{GeV}$.

3rd scenario: for even smaller $\kappa_L^{\text{MFV}}$, both the three-body and two-body decays are mediated \emph{indirectly}, namely when $\kappa_L^{\text{MFV}}\ll \varepsilon\, N_8 \mathcal{C}_i$.
This scenario is uniquely realized for low UV scales when $\tilde{c}_{BB}(\Lambda)$ is the only non-vanishing coupling, in which case we recover the SM scaling,
\begin{align}\label{eq:RMFV3}
    R_0,R_{\pm} \sim \frac{1}{|\epsilon|^2}\,I_{0,\pm}(m_a) \sim \frac{\text{Br}[K_L \to 3\pi^0]}{\text{Br}[K_L \to 2\pi^0]}\,,
\end{align}
where $1/|\varepsilon|^2 \approx 2 \times 10^5$.
This is the largest possible enhancement due to the existing CP-violation within the SM.
It is important to note that since the indirectly-mediated three-body decay requires some non-vanishing coupling to the strong sector, having either $\tilde{c}_{WW}(\Lambda)$ or $\tilde{c}_{BB}(\Lambda)$ as the only non-vanishing coupling requires some non-trivial cancellations to take place; more details are provided in \App{app:MFV_details}.
We summarize the results of this section in \Tab{tab:MFV_summary}.
\begin{table}   
\begin{center}
\begin{tabular}{ |c |c| c| }
\hline
MFV scenario& $R_{0,\pm}/I_{0,\pm}(m_a)$ & Realization  \\
\hline
$\kappa_L^{\text{MFV}}\gg N_8$ (direct)&  $\left(\frac{\text{Re}\, V^*_{td}}{\text{Im}\,V^*_{td}} \right)^2 \sim 10$ & $c_{L}^{\text{\tiny NP}} \neq 0$ or $c_{tt}(\Lambda) \neq 0$  \\
\hline
& \quad \quad \quad \quad \quad \quad \quad \quad  $20-600$ & $\tilde{c}_{GG}(\Lambda) \neq 0$  \\
$N_8 \gg \kappa_L^{\text{MFV}}\gg \varepsilon\, N_8$ (mix)& $\left(\frac{N_8}{\text{Im}\,\kappa_L^{\text{MFV}}} \right)^2 \sim 1600 $   & $\tilde{c}_{WW}(\Lambda) \neq 0$  \\
& \quad \quad \quad \quad \quad \quad \quad \quad  $1.6\times 10^5$  & $\tilde{c}_{BB}(\Lambda) \neq 0 \;\;(\Lambda \gg 10^6\,\text{GeV})$  \\
\hline
$\varepsilon\,N_8 \gg \kappa_L^{\text{MFV}} $ (indirect)&  $|\varepsilon|^{-2} \sim 2\times 10^5$ & $\tilde{c}_{BB}(\Lambda) \neq 0 \;\;(\Lambda \ll 10^6\,\text{GeV})$  \\
\hline
\end{tabular}
\end{center}
\caption{
Summary of the 3 possible hierarchies in the MFV scenarios.
For more details, see main text and \App{app:MFV_details}.}
\label{tab:MFV_summary}
\end{table}
\begin{figure}[t]
  \centering   \includegraphics[width=0.49\textwidth]{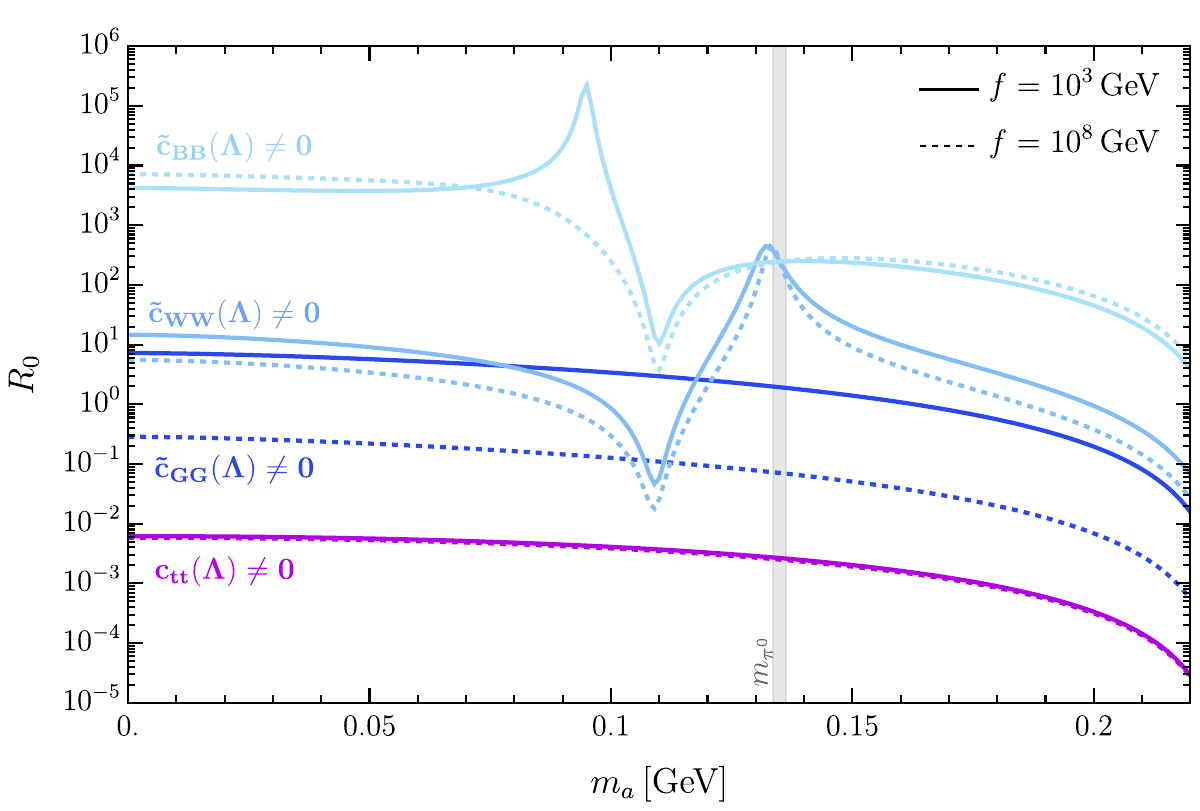}  
\includegraphics[width=0.49\textwidth]{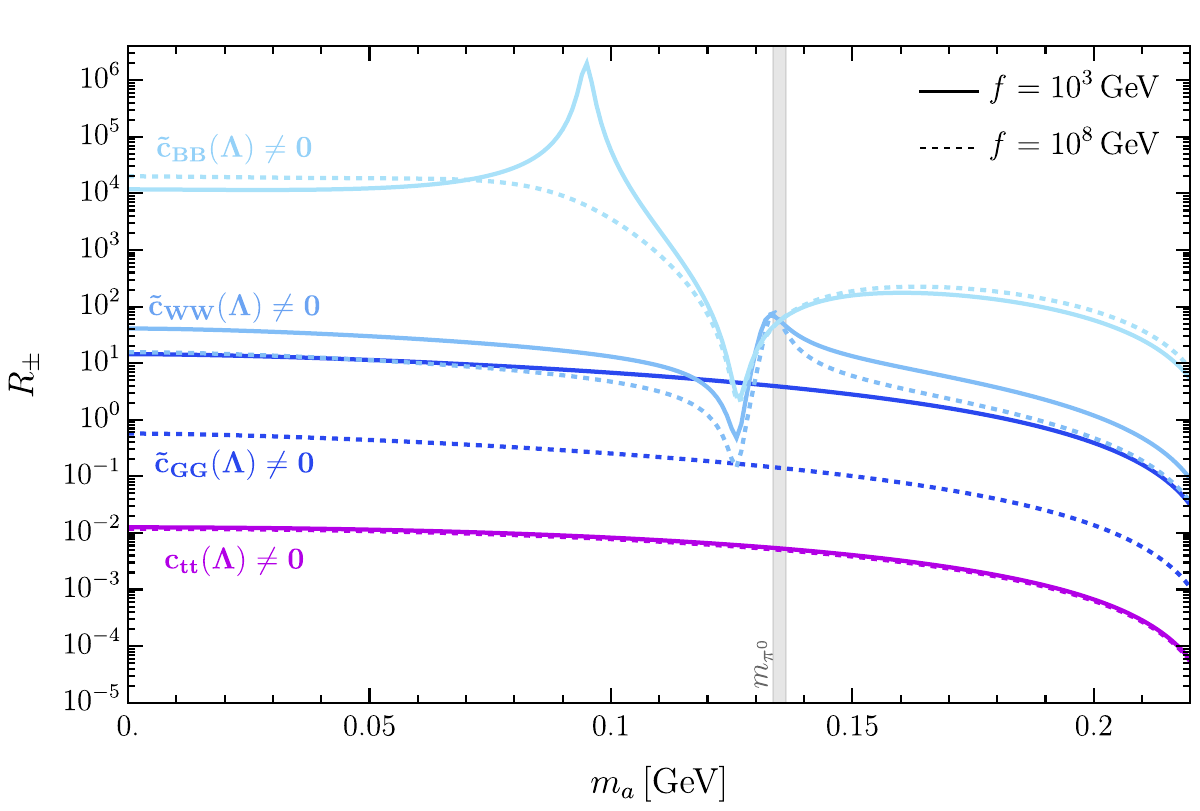}   
      \caption{The ratios $R_0$ (left panel) and $R_{\pm}$ (right panel) defined in Eq. (\ref{eq:ratio_defs}), as a function of the ALP mass, $m_a$, for the various MFV scenarios discussed in the text, plotted in solid and dashed lines for $f=\Lambda/4\pi = 10^3\,$GeV and $ 10^8\,$GeV, respectively.   }     \label{fig:ratios_numerics}
\end{figure}
In \Fig{fig:ratios_numerics} we plot the numerical results for $R_0$ and $R_{\pm}$ as a function of the ALP mass $m_a$ for several scenarios discussed in the text.
We find good agreement with our estimates for $m_a \ll m_{\pi}$. The dependence of the $\mathcal{C}_i$ coefficient on the ALP mass leads to a non-trivial behavior for heavier ALP masses.
\subsection{Grossman-Nir bound}
\label{sec:GN_bound}
After establishing that in some theories the three-body decay can be the dominant channel for long-lived neutral kaons, it is useful to compare the three-body rate to the charged kaon decay rate.
Since the latter does not require any CP violation, the charged kaon decay usually provides a stronger probe of FV and can be used to place an upper bound on the neutral decay rates using the Grossmann-Nir (GN) bound~\cite{Grossman:1997sk}.
We are interested in the ratios,
\begin{align}
   \frac{\Gamma(K_L \to \pi^0 \pi^0 a)}{\Gamma(K^+ \to \pi^+  a)}\,, \;\;  \frac{\Gamma(K_L \to \pi^+ \pi^- a)}{\Gamma(K^+ \to \pi^+  a)}.
\end{align}
These ratios differ from $\Gamma(K_L \to \pi^0\,a)/\Gamma(K^+\to \pi^+\,a)$ used in the derivation of the GN bound, since the process in the numerator  is (1) $CP$ preserving and (2) involves three particles in the final state.
We expect these ratios to be consistently much smaller than unity because of the phase-space suppression.

However, as the ALP mass approaches the pion mass, large deviations from this expectation can occur due to resonance enhancement. 
The $\mathcal{O}(N_8)$ contributions to $K_L\to \pi^0\pi^0a$ diverges as $m_a \to m_{\pi^0}$, while a similar contribution to $K^+\to \pi^+a$ remains finite. 
The divergence signals the breakdown of the small-angle approximation used in the calculation, which schematically holds as long as
\begin{align}
   \theta_{\pi^0a}\sim \left|\frac{m^2_{\pi^0}}{m_a^2-m^2_{\pi^0} }\cdot \frac{f_\pi}{f}  \right|\ll 1\,,
\end{align}
or equivalently in terms of a tuning parameter $\delta_{m_a} \equiv {m_a^2}/{m_\pi^2}-1$,
\begin{align}
    f_\pi/f \ll |\delta_{m_a}|\,.
\end{align}
This divergence can be avoided by properly diagonalizing the ALP-pion system, in which case the mixing angle tends to $\pi/4$ as the masses become degenerate, and the divergence is regulated by an $\arctan$ function.
However, in practice, for most reasonable values of $f$ the small-angle approximation is still valid.
Even for a moderately low UV scale, e.g. $f=100\,\text{GeV}$, a percent-level mass tuning $\delta_{m_a}\sim0.01$ falls well within the range of the small-angle approximation.
A large mixing angle could in principle induce observable modifications to neutral pion properties.
We defer this investigation to future work.

\begin{figure}[t]
  \centering    
\includegraphics[width=0.95\textwidth]{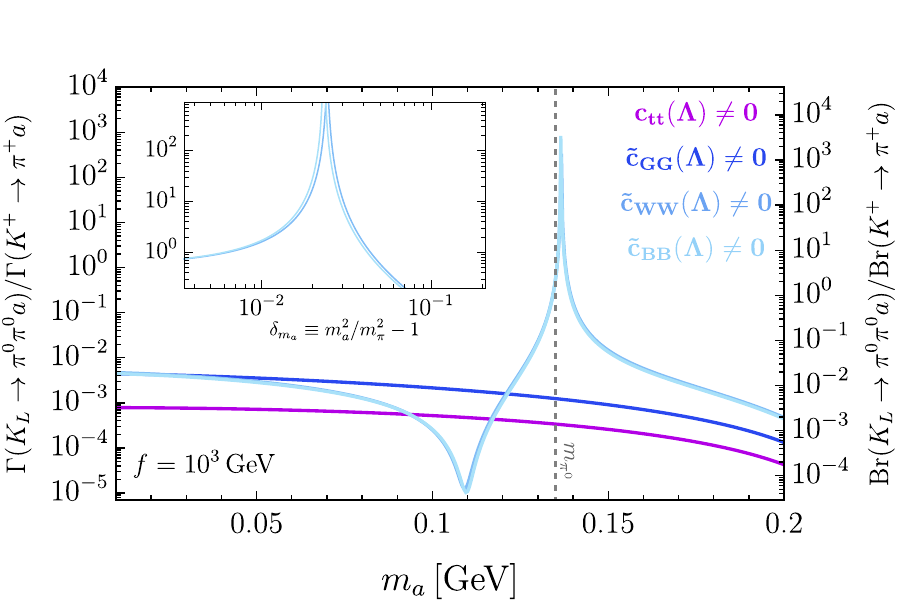}
      \caption{The ratio $\Gamma(K_L \to \pi^0 \pi^0 a)/\Gamma(K^+ \to \pi^+  a)$ as a function of ALP mass $m_a$ for the various MFV cases discussed in the text. The ratio is calculated for fixed $f=10^3\,$GeV.
      The ratio is visibly enhanced around the pion mass for $\tilde{c}_{WW}\neq0$ or $\tilde{c}_{BB}\neq0$. The insert plot shows the enhancement as a function of the tuning of the ALP mass around the pion mass.
      For reference we plot the corresponding values of $\text{Br}(K_L \to \pi^0 \pi^0 a)/\text{Br}(K^+ \to \pi^+ a)$ on the $y$ axis on the RHS.}     \label{fig:GN}
\end{figure}

In \Fig{fig:GN} we plot the ratio $\Gamma(K_L \to \pi^0 \pi^0 a)/\Gamma(K^+ \to \pi^+  a)$ for the different MFV scenarios as a function of the ALP mass for fixed $f=10^3\,$GeV.
For this value of $f$, the small-angle approximation is valid for mass-tuning $\delta_{m_a} \gg 10^{-4}$.
We first note that the $c_{tt}\neq0$ and $c_{GG}\neq0$ scenarios do not display any deviation from the expected phase-space suppression.
This is due to the dominance of the indirect contribution $\kappa_V \gg N_8$. 
In these models, only when the ALP mass is tuned to a very large degree, $\delta_{m_a} \sim 10^{-9}$, do the resonance enhancement effects become relevant.
Such tuning is questionable from a theoretical point of view, and depending on the value of $f$ may also fall outside the validity of the small angle approximation.

The remaining $c_{WW}\neq0$ and $c_{BB}\neq0$ scenarios are dominated by the direct weak-interaction contribution. In this case, to reliably compute the charged kaon rate close to the pion mass we must add the additional weak-interaction contribution from the $\mathcal{O}_{27}^{3/2}$ %and $\mathcal{O}_{27}^{1/2}$ 
operator~\cite{Bauer:2020jbp,Bauer:2021mvw}. This contribution is
usually safely neglected as it appears with a numerical coefficient smaller than $N_8$ by a factor of $\sim 30$%and $\sim 150$, respectively
~\cite{Neubert:1991zd}. 
However, near the pion mass, the $\mathcal{O}_{27}^{3/2}$ contribution to $K^+\to \pi^+a$ becomes resonantly enhanced and can therefore dominate the contribution from the octet operator. The full rate can be written as%, and must therefore be included in the calculation as it can dominate over the finite $\mathcal{O}(N_8)$ contribution 
\begin{align}
    \mathcal{M}(K^+ \to \pi^+ a) =& \frac{i(m_K^2-m_\pi^2)}{2f} \bigg(    \kappa_V+N_8[\mathcal{C}_1+\Delta \mathcal{C}_1]+N_{27}^{3/2} \mathcal{C}_{27}\bigg),
    \label{eq::charged_O27}
\end{align}
with $|N_{27}^{3/2}| \equiv |G_F V_{us}V^*_{ud}g_{27}^{3/2} f_\pi^2/\sqrt{2}| = 4.86\times 10^{-9}$~\cite{Neubert:1991zd}.
Our calculation of $\mathcal{C}_{27}$
is consistent with previous results~\cite{Bauer:2021mvw,Bauer:2020jbp}, with the full expression given in \App{app:Ci}.
 
 \Fig{fig:GN} shows that the ratio $\Gamma(K_L\to\pi^0\pi^0a)/\Gamma(K^+\to\pi^+a)$ spikes near the pion mass, $\delta_{m_a} \sim\mathcal O( 10^{-2})$, a tuning which is within the validity of the small-angle approximation for our chosen value of $f$.
This can be understood via expanding the charged decay amplitude in \Eq{eq::charged_O27} around $m_a=m_{\pi^0}$, and working to leading order in $m_{\pi^0}/m_K$
\begin{equation}
     \mathcal{M} \simeq \frac{im_K^2}{2f}\left[ \,\frac{3N_{27}^{3/2} \left(c_u-c_d\right)}{2 \delta_{m_a} } -\frac{N_8}{2} \left( c_d+c_s+2 c_u+4 c_{GG}+\Delta k_{V}^{d-s}\right)\right]\,,
\end{equation}
where we used the notation (see also \App{app:Ci})
\begin{align}
        c_{q} &\equiv k_{q}-k_{Q}\,, \;\;\;\; \Delta k_{V}^{d-s}\equiv k_{d}+k_{D}-(k_{s}+k_{S})\,.
\end{align}
The cancellation occurs when
\begin{align}
  \delta_{m_a} = \frac{N^{3/2}_{27}}{N_8}\frac{3 \left(c_u-c_d\right)}{ \left( c_d+c_s+2 c_u+4 c_{GG}+\Delta k_{V}^{d-s}\right)  } \approx \mathcal{O}\left( \frac{N^{3/2}_{27}}{N_8} \right) \approx \mathcal{O}\left( 10^{-2} \right)\,, 
\end{align}
consistent with our numerical results (see the inset in the
figure).

%%%%%%%%%%%%%%%%%%%%%%
%%%%%%%%%%%%%%%%%%%%%%
\section{Phenomenology}
\label{sec:pheno}
%%%%%%%%%%%%%%%%%%%%%%
%%%%%%%%%%%%%%%%%%%%%%
\subsection{Long-lived ALPs}
\begin{figure}[t]
  \centering   \includegraphics[width=0.8\textwidth]{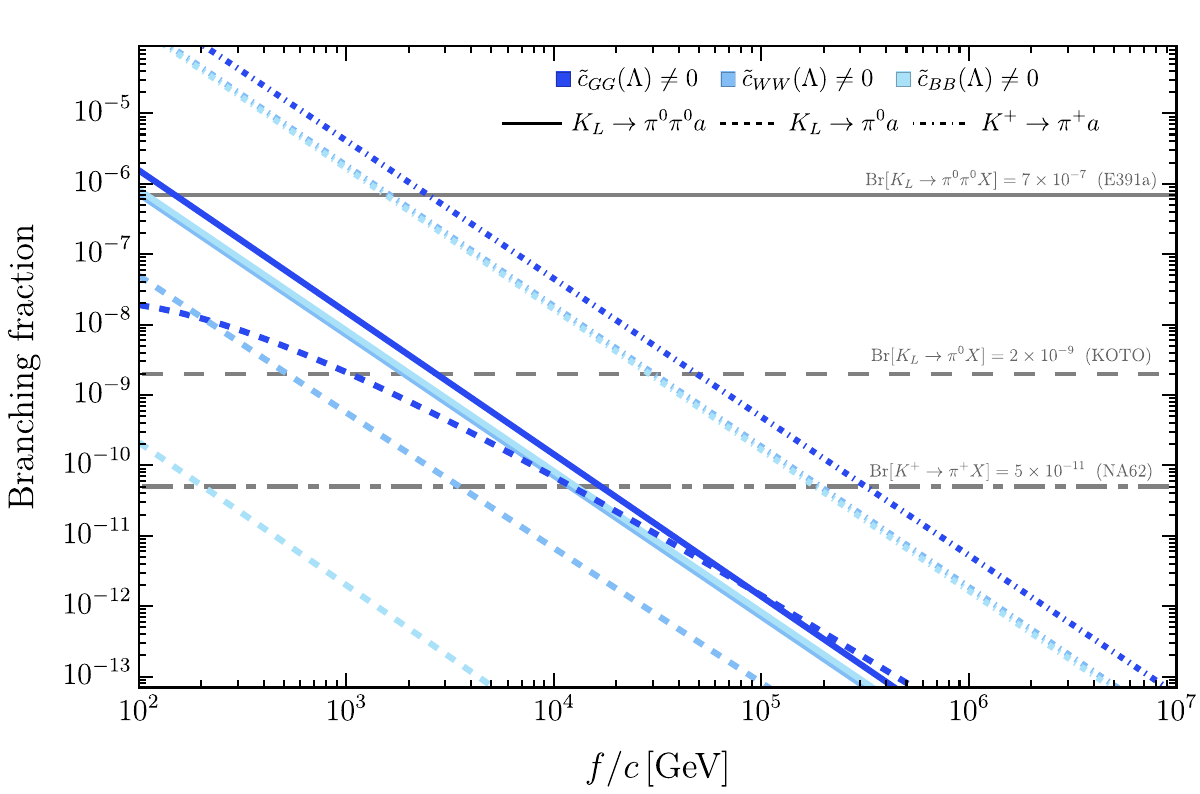}  
      \caption{Branching ratio of $K_L \to \pi^0\pi^0 \,a$, $K_L \to \pi^0 \,a$, and $K^+ \to \pi^+ \,a$ in solid, dashed and dot-dashed, respectively, as a function of the effective scale $f/c$ for $m_a=0.01\,$GeV.
      $c$ is a single Wilson coefficient needed to realize the different scenarios, see \App{app:MFV_details} for more details.
      In different colors we plot three scenarios in which the three-body neutral kaon decay is larger than the two-body one. 
      The gray horizontal lines represent the relevant experimental bounds.}     \label{fig:numeric_rates}
\end{figure}

%%%%%%%%%%%%%%%%%%%%%%
%%%%%%%%%%%%%%%%%%%%%%
The results of the previous section can have interesting phenomenological consequences in the context of experimental searches.
In regions of parameter space where the ALP is long-lived and escapes detection, i.e. $m_a \lesssim 0.1\,$GeV, the strongest existing bounds on kaon decays are
\begin{align}\label{eq:NA62}
    \text{Br}[K^+ \to \pi^+ \,X] &< (3-6) \times 10^{-11}\; &&(\text{NA62~\cite{NA62:2021zjw}})\,,&&
    \\
    \text{Br}[K_L \to \pi^0\, X] &< 1.6 \times 10^{-9} \;&&(\text{KOTO~\cite{KOTO:2024zbl}})\,, \label{eq:K_pi0a_bound}
        \\
    \text{Br}[K_L \to \pi^0\,\pi^0\, X] &< 7 \times 10^{-7} \;&&(\text{E391a~\cite{E391a:2011aa}})\,,\label{eq:K_pi0pi0a_bound}
\end{align}
where the range in Eq. (\ref{eq:NA62}) corresponds to $X$ masses in the range $0-110\,\text{MeV}$. 
In \Fig{fig:numeric_rates}, we plot the $K^+$ and $K_L$ decay rates for several scenarios in which the $K_L$ three-body rate is larger than the $K_L$ two-body rate.
The charged kaon decays constraints $f/c\gtrsim 10^5\,$GeV  (NA62 bound in  Eq. (\ref{eq:NA62})), while the constraint due to neutral three-body decays is significantly weaker,  $f/c\gtrsim 10^2\,$GeV (E391a bound in  Eq. (\ref{eq:K_pi0pi0a_bound})), where $c$ is a single Wilson coefficient needed to realize the different scenarios, see \App{app:MFV_details} for more details.
In these types of searches, the charged kaon decay typically leads to a more stringent bound, since the rate is insensitive to the $CP$ properties of the flavor-violating coupling.
In addition, it benefits from an enhancement due to the larger available phase space, see \Sec{sec:GN_bound}. 
In these types of models, an improved bound on $\text{Br}[K_L \to \pi^0\,\pi^0\, X]$, which is currently the weakest of the three, could lead to significantly stronger constraint compared to the constraint coming from $K_L \to \pi^0\, X$.
In order to be competitive with bounds from charged kaon decay, the bound on $\text{Br}[K_L \to \pi^0\pi^0\,X]$ should be $\mathcal{O}(10^3)$ stronger than the bound on $\text{Br}[K^+ \to \pi^+\,X]$ for $m_a \lesssim 0.1\,\text{GeV}$, namely around $10^{-13}$, see \Fig{fig:numeric_rates}.
However, as long as the ALP remains sufficiently long-lived, for ALPs with masses closer to the pion mass that requirement gets weaker as $K_L \to \pi^0\pi^0\,X$ becomes comparable or even dominant over the charged decay in some models, see \Fig{fig:GN}.  
There is currently no direct bound on $\text{Br}[K_L \to \pi^+\,\pi^-\, X]$.
The different experimental signature in such a search could be potentially cleaner due to the stability of charged pions compared to their neutral counterpart.  
%%%%%%%%%%%%%%%%%%%%%%
%%%%%%%%%%%%%%%%%%%%%%
%%%%%%%%%%%%%%%%%%%%%%
\subsection{Promptly-decaying ALPs}
\label{section:prompt_ALP}
\begin{figure}[t]
  \centering   \includegraphics[width=0.95\textwidth]{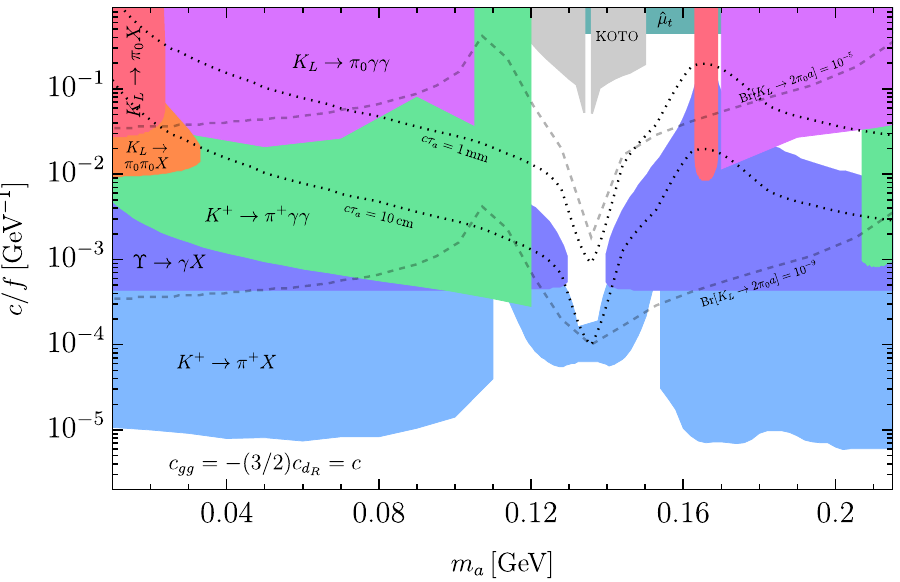}  
      \caption{Experimental constraints on the $\tilde{c}_{BB}(\Lambda)\neq0$ scenario in the $\{m_a,c/f \}$ plane, realized by taking $c_{gg}=-(3/2)c_{d_R}\equiv c$.  
      References for the experimental constraints shown in color are given in the main text.
      The gray region shows our estimate of the exclusion limit derived using KOTO calibration data~\cite{Balkin:2025aer}.
      To illustrate the phenomenology, we plot as gray dotted curves contours of constant lifetime, while the gray dashed curves are  contours of fixed $\text{Br}[K_L \to \pi^0 \pi^0 a]=\{10^{-5},10^{-9}\}.$}     
      \label{fig:K_bounds}
\end{figure}

%%%%%%%%%%%%%%%%%%%%%%
%%%%%%%%%%%%%%%%%%%%%%
For $m_a \gtrsim 0.1\,$GeV, the ALP can decay promptly, leaving a visible signature.
We denote the low-energy ALP coupling to  photons by  $C_{\gamma\gamma}^{\rm eff}$, such that the decay width is then $\Gamma(a \to \gamma\gamma)
\equiv
 \alpha^2 m_a^3\left| C_{\gamma\gamma}^{\rm eff} \right|^2/(64 \pi^3 f^2)
$.
For more details on $C_{\gamma\gamma}^{\rm eff}$, see \App{app:photon_coupling}. 
In \Fig{fig:K_bounds} we show the experimental bounds in the $\{m_a,c/f\}$ plane for the benchmark model $\tilde{c}_{BB}(\Lambda)\neq0$, which can be realized by taking $c_{GG} = -(3/2)c_{d_R}\equiv c$ as the only non-vanishing couplings in the UV theory.
In this benchmark model, the ALP decays predominantly to photons, i.e. $\text{Br}[a \to \gamma\gamma]\approx 1$, due to its coupling to gluons.
We plot the experimental bounds for both an invisible ALP and an ALP decaying into photons: $K^+ \to \pi^+ X$~\cite{NA62:2021zjw}, $K^+ \to \pi^+ \gamma \gamma$~\cite{NA62:2023olg,E949:2005qiy}, $K_L \to \pi^0 X$~\cite{KOTO:2024zbl}, $K_L \to \pi^0 \gamma \gamma$~\cite{NA48:2002xke} and $K_L \to \pi^0 \pi^0 X$~\cite{E391a:2011aa}.
For completeness we also include the bounds on the top chromomagnetic dipole moment $\hat{\mu}_t$~\cite{CMS:2019nrx} and the $\Upsilon(1S) \to \gamma X$~\cite{BaBar:2010eww}, using the expressions given in Ref.~\cite{Bauer:2021mvw}.\footnote{
In principle,  measurements of the $K^+$ branching ratios to SM could also provide an additional bound on NP contributions to its total width~\cite{Goudzovski:2022vbt}.
However, interpreting these measurements in terms of a NP contribution is non-trivial around the pion mass, where the ALP could be misidentified as a pion, and is beyond the scope of this work.
Away from the pion mass, other experiments provide stronger bounds. 
We therefore we do not consider this bound here.}
For the searches performed in~\cite{E949:2005qiy,KOTO:2024zbl,NA48:2002xke,E391a:2011aa,BaBar:2010eww}, we perform simple recasts to account for the finite lifetime of the ALP. The details are provided in \App{app:exp_recasts}.
As expected, the strongest constraints are due to the charged kaon decays both in the invisible as well as in prompt searches.
It is interesting to note that although the experimental bound on $K_L \to \pi^0 X$ is more than two orders of magnitude stronger than $K_L \to \pi^0 \pi^0  X$ (see \Eq{eq:K_pi0a_bound} and \Eq{eq:K_pi0pi0a_bound}), the latter provides a stronger experimental bound due to the suppressed two-body decay rate (see red vs. orange regions in the figure).  
Interestingly, the neutral three-body decay could be used to probe the unexplored region around the pion mass, as we demonstrated in Ref.~\cite{Balkin:2025aer} by using KOTO's calibration data for $K_L \to 3\pi^0 \to 6\gamma$.
Our estimate of the resulting experimental bound is shown in gray in \Fig{fig:K_bounds}. 

%%%%%%%%%%%%%%%%%%%%%%
%%%%%%%%%%%%%%%%%%%%%%

%%%%%%%%%%%%%%%%%%%%%%
\section{Conclusions}
\label{sec:conc}
We presented a detailed analysis of three-body decays of the neutral kaon involving axion-like particles (ALPs). 
We emphasize the importance of the weak-interaction contribution arising from the interplay of SM weak currents and flavor-preserving ALP–quark couplings.
Depending on the UV completion, this contribution can dominate the flavor violation driving three-body decays. 
This includes the well-motivated case of an ALP that exclusively couples to gluons.

While charged kaon decays strongly probe flavor violation, neutral kaon decays uniquely probe both the $CP$ properties and flavor structure of ALP couplings.
We compared the three-body decay $K_L \to \pi \pi\,a$, which requires flavor violation but not $CP$ violation, to the two-body decay $K_L \to \pi^0\,a$, which requires both.
Interestingly, ALP couplings in models which respect the MFV hypothesis do not introduce new CPV phases at leading order in the MFV expansion, leading to sharp predictions for CPV observables in such models.
Conversely, observing a large CPV phase would be inconsistent with the MFV hypothesis.

Unlike in two-body decays, we found that the three-body calculation required the inclusion of naively factorizable terms to cancel unphysical contact terms.
With our basis-independent results - including both direct and indirect weak contributions - we identified classes of models in which $CP$ violation from UV couplings is suppressed.
In such models, ALP production via the three-body decay can be comparable to, or even dominate over, the two-body decay, despite the expected phase-space suppression. 

Finally, we discuss some of the phenomenological implications of our results.
Although charged kaon decays impose the strongest constraints on much of the parameter space, neutral two- and three-body kaon decays can in principle probe unexplored regions around the pion mass at low values of $c/\Lambda$.
Indeed, we found that the charged and neutral decay rates become comparable when the ALP is sufficiently degenerate with the pion, thus strongly violating the Grossman-Nir bound.
We leave a detailed exploration of current and future experimental sensitivity to future work.
It would also be interesting to investigate the three-body decay 
$K_L \to \pi^+ \pi^- a$, which currently lacks direct experimental constraints.
This channel could offer a cleaner experimental signature due to the stability of charged pions compared to their neutral counterparts.

\acknowledgments
We thank Y. Grossman for early discussions on the topic. 
We thank D. J. Robinson for collaboration in the early stages of this work and for useful feedback on the manuscript.
The research of RB and SG is supported in part by the U.S. Department of Energy grant number DE-SC0010107. 
This work was performed in part at the Aspen Center for Physics, which is supported by National Science Foundation grant PHY-2210452. 
CS is supported by the Office of High Energy Physics of the U.S. Department of Energy under contract DE-AC02-05CH11231.

%%%%%%%%%%%%%%%%%%%%%%
\appendix
\section{Chiral perturbation theory}
\label{app:chi_PT}
In this appendix we provide additional details on the calculation of the kaon decay amplitudes in chiral perturbation theory.
Following the approach of Ref.~\cite{Bauer:2021wjo},
before matching the theory to the low-energy chiral Lagrangian, it is convenient to perform a generic field redefinition, as specified in Sec. \ref{Sec:chipT}
\begin{align}
q(x) \to \text{exp}\left[-i(\boldsymbol{\delta}_q + \boldsymbol{a}_q \gamma_5 )c_{GG} \frac{a(x)}{f} \right]q(x)\,,
\end{align}
where $\boldsymbol{\delta}_q$ and $\boldsymbol{a}_q$ are diagonal matrices in flavor space.
This field redefinition removes the gluon coupling if $\text{Tr}[\boldsymbol{a}_q] =1$. 
Importantly, any physical observable must not depend on the base-dependent parameters, $\boldsymbol{\delta}_q$ and $\boldsymbol{a}_q$.
The field redefinition shifts and rotates the ALP coupling to quarks, 
\begin{align}
&{\bf k}_Q \to \hat{\bf k}_Q(a) = U_-({\bf k}_Q+\phi_q^-)U^\dagger_-\,,
\\
&{\bf k}_q \to \hat{\bf k}_q(a) = U_+({\bf k}_q+\phi_q^+)U^\dagger_+\,,
\end{align}
where $\phi_q^\pm \equiv c_{GG} (\boldsymbol{\delta}_q \pm \boldsymbol{a}_q )$ and $U_\pm(a) \equiv e^{i \phi_q^\pm a/f}$. The field redefinition also leads to the axion-dressed quark mass
\begin{align}
{\bf m}_q \to \hat{\bf{m}}_q =  \text{exp}\left[-2i \boldsymbol{a}_q c_{GG} \frac{a(x)}{f} \right]{\bf m}_q\,,
\end{align}
where ${\bf m}_q = \text{Diag}(m_u,m_d,m_s)$.
The UV theory is matched to the chiral Lagrangian,
\begin{align}
\mathcal{L}_{\chi} = \mathcal{L}_{\text{\tiny kin}}+\mathcal{L}_{ m}+\mathcal{L}_{\text{\tiny weak}}\,,
\end{align}
with
\begin{align}
\mathcal{L}_{\text{\tiny kin}} &=  \frac12 (\partial_\mu a)^2+\frac{f_\pi^2}{8}\text{Tr}[D^\mu \Sigma (D_\mu \Sigma)^\dagger]\,,
\\
\mathcal{L}_{ m} &= \frac{f_\pi^2}{4} B_0 \text{Tr} [\hat{\bf m}_q \Sigma^\dagger+\text{h.c}]-\frac12 m_a^2 a^2\,,
\\
\mathcal{L}_{\text{\tiny weak}} &= \frac{4 N_8}{f_\pi^2}[L_\mu L^\mu]^{32}\nonumber\\
&+\frac{4 N^{3/2}_{27}}{f_\pi^2}\left[(L_\mu)_{32} (L^\mu)_{11}+(L_\mu)_{31} (L^\mu)_{12}-(L_\mu)_{32} (L^\mu)_{22}\right]+\text{h.c}\,,
\end{align}
with $B_0=m_\pi^2/(m_u+m_d)$  and where we defined
\begin{align}
 D_\mu \Sigma &\equiv \partial_\mu \Sigma-i \frac{\partial_\mu a}{f} (\hat{\bf k}_Q\Sigma-\Sigma \hat{\bf k}_q)\,,
 \\
 L_\mu^{ji} &\equiv -\frac{i f^2_\pi} {4}e^{i(\phi_{q_i}^--\phi_{q_j}^-)a(x)/f}[\Sigma (D_\mu \Sigma)^\dagger]^{ji}
 \\
 N_8 &\equiv -\frac{G_F}{\sqrt{2}} V_{us}V^*_{ud}g_8 f_\pi^2\,, \;\;\; |N_8| \approx 1.53\times10^{-7}\,,
  \\
 N_{27}^{3/2} &\equiv -\frac{G_F}{\sqrt{2}} V_{us}V^*_{ud}g_{27}^{3/2} f_\pi^2\,, \;\;\; |N_{27}^{3/2}| \approx 4.86\times 10^{-9}\,,
 \end{align}
with $G_F$ the Fermi constant, $G_F=1.1663787\times 10^{-5}$ GeV$^{-2}$ and $|g_8|\sim 5$.
 The field $\Sigma$ is given by $\Sigma=e^{2i \Pi/f_\pi}$ with
\begin{align}
\Pi = 
\begin{pmatrix}
\dfrac{\eta}{\sqrt{6}} + \dfrac{\pi^{0}}{\sqrt{2}} & \pi^{+} & K^{+} \\[8pt]
\pi^{-} & \dfrac{\eta}{\sqrt{6}} - \dfrac{\pi^{0}}{\sqrt{2}} & K^{0} \\[8pt]
K^{-} & \bar{K}^{0} & -\sqrt{\dfrac{2}{3}}\,\eta
\end{pmatrix}\,.
\end{align}

\section{Phase space integrals and ratios}
\label{app:PSI_and_ratios}
As we discussed in Secs. \ref{Sec:Kpi0pi0} and \ref{Sec:Kpi+pi-}, the amplitude of the three-body decays can be written as a function of the two kinematical variables, $s_\pi $ and 
$\tilde{s}$. The phase space ratios defined in Eqs. (\ref{eq:ratio_schematic}), (\ref{Eq:I}) can be written as
\begin{align}
    I_{\pm}(m_a,\{a_1,a_2,a_3\}) &\equiv \frac{\frac12\int \mathrm{d}\Pi_3\left|        a_1+a_2\frac{s_\pi}{m_K^2}+ a_3\frac{\tilde{s}}{m_K^2}\right|^2}{(|a_1|^2+|a_2|^2+|a_3|^2)(1-m_\pi^2/m_K^2)^{2} f_\pi^2\int \mathrm{d}\Pi_2}\,,
    \\
    I_{0}(m_a,\{a_1,a_2\}) &\equiv \frac12 I_{\pm}(m_a,\{a_1,a_2,a_3=0\}),
\end{align}
where $a_1,a_2$ and $a_3$ are coefficients which depend on the amplitude of the decay mode under consideration and the $f_\pi^2$ normalization is used to make the $I_{0,\pm}$ dimensionless.

The two-body phase space integral is instead a simple function of $m_a$,
\begin{align}
    \int \mathrm{d}\Pi_2 = \frac{\sqrt{1-(m_a+m_\pi)^2/m_K^2} \sqrt{1-(m_a-m_\pi)^2/m_K^2}}{8 \pi}\,.
\end{align}
The three-body phase space depends on the kinematic variables $s_\pi$ and $s_{\pi a}\equiv (p_a+p_{\pi_1})^2$ (where $\tilde{s} = 2 s_{\pi a}+s_{\pi}-m_{a}^2-m_K^2-2 m_\pi^2$) and
\begin{align}
  \mathrm{d}\Pi_3 = \frac{ \mathrm{d}s_\pi \mathrm{d}s_{\pi a}}{ 128 \pi^3 m_K^2}\,.
\end{align}
The integral can be solved numerically, using the $s_\pi$-dependent integration range for $s_{\pi a}$
\begin{align}
    &({s}_{\pi a})_\pm
    &= \frac{1}{2} \left(m_a^2+m_K^2+2 m_\pi^2-s_\pi\pm\sqrt{s_\pi-4 m_\pi^2} \sqrt{\frac{\left(m_a^2-m_K^2\right)^2}{s_\pi}-2 \left(m_a^2+m_K^2\right)+s_\pi}\right)\,,
\end{align}
and for $s_\pi$,
\begin{align}
  (2 m_\pi)^2  <s_\pi< (m_K-m_a)^2\,.
\end{align}

\section{Maximal flavor violation}
\label{sec:non_MFV}
In maximally flavor violating models, the flavor symmetry breaking parameter in the UV theory is much larger than the SM flavor symmetry breaking sources,
\begin{align}
    \kappa^{\text{\tiny FV}} \gg \text{Max}[\kappa_L^{\text{\tiny MFV}},N_8\, \mathcal{C}_i]\,.
\end{align}
For clarity, we introduce the flavor-blind parameters, $\mathcal{C}_i$, which were mentioned in the main text and whose explicit forms are reported in App. \ref{app:Ci}. 
As a consequence, the three-body decay is dominated by this coupling projected in the $CP$ conserving direction, while
the two-body decay is dominated by this coupling projected in the $CP$-breaking direction.
Thus, the ratio of rates strongly depends on the phase of the coupling, which we denote as $\theta_\kappa \equiv \text{arg}\,\kappa^{\text{FV}}$. 
In this section, we will explain the scaling in Eqs. (\ref{eq:maximallyR0}), (\ref{eq:maximallyRpm}).

\subsection{\texorpdfstring{$\tan \theta_\kappa \sim \mathcal{O}(1)$}{tan theta ~ O(1)}}
\label{sec:Non_MFV_case_I}
In this case, the UV coupling maximally violates not only the flavor but also the $CP$ symmetry. 
We can neglect all the contribution proportional to $N_8$ and $\varepsilon$, to find the ratios defined in Eq. (\ref{eq:ratio_defs})
\begin{align}
\label{eq:rates_neutral}
    R_{0}(m_a) &= (c_{\theta_{\kappa}}/s_{\theta_{\kappa}})^2\left(1 + \frac{1}{(1-x_a)^2}\right){I}_0\left(m_a,\left\{-1,\frac{1}{1-x_a}\right\}\right)\,,
    \\
    \label{eq:rates_charged}
    R_{\pm}(m_a) &= (1/s_{\theta_{\kappa}})^2 \left(c_{\theta_{\kappa}}^2 + \frac{1}{(1-x_a)^2}\right){I}_\pm\left(m_a,\left\{-c_{\theta_{\kappa}},\frac{c_{\theta_{\kappa}}}{1-x_a},\frac{-i\, s_{\theta_{\kappa}}}{1-x_a}\right\}\right)\,,
\end{align}
where $x_a \equiv m_{a}^2/m_K^2$, $c_\alpha \equiv \cos\alpha$, and $s_\alpha \equiv \sin\alpha$.
The ratios of phase space integrals $I_0,I_{\pm}$ are defined in \App{app:PSI_and_ratios}.
These functions can be calculated numerically.
For $\theta_\kappa \sim \mathcal{O}(1)$, the ratios are determined by the phase space ratio, since the prefactor is of $\mathcal O(1)$. These monotonically decreasing functions are bounded from above
\begin{align}
 &{I}_0 \leq {I}_0\left(0,\left\{-1,1\right\}\right)\approx 4\times 10^{-4}\,,
 \label{eq:ratios_maximum_CP_violation_neutral}
    \\
   &{I}_\pm \leq {I}_\pm\left(0,\left\{-1,1,0\right\}\right) \approx 8\times 10^{-4},  \label{eq:ratios_maximum_CP_violation_charged}
\end{align}
and vanish as $m_a \to m_K-2m_\pi$.
This is the typical phase space suppression encountered when comparing two-body and three-body decays.
Note that for the charged decay, the phase space integral depends weakly on $\theta_\kappa$, e.g. for $m_a=0$ the functions is minimized for $\theta_\kappa = \pi/2$, where $I_\pm \approx 2\times 10^{-4}$.
\subsection{\texorpdfstring{$\tan \theta_\kappa \ll |\varepsilon|$}{tan theta << eps}}
In this case, $\kappa^{\text{\tiny FV}}$ is a weaker source for $CP$ violation compared to the SM $\varepsilon$ parameter, and the latter is the dominant contribution to the two-body decay.  
Here, we have a similar expressions as in the previous subsection, but with a larger coupling hierarchy,
\begin{align}
    R_{0}(m_a)  &= \frac{1}{|\varepsilon|^2}\left(1 + \frac{1}{(1-x_a)^2}\right){I}_0\left(m_a,\left\{-1,\frac{1}{1-x_a}\right\}\right)\,,
    \\
    R_{\pm}(m_a) &= \frac{1}{|\varepsilon|^2}\left(1 + \frac{1+|\varepsilon|^2}{(1-x_a)^2}\right){I}_{\pm}\left(m_a,\left\{-c_{\theta_{\kappa}},\frac{c_{\theta_{\kappa}}}{1-x_a},\frac{-\varepsilon c_{\theta_{\kappa}}}{1-x_a}\right\}\right)\approx 2R_0(m_a)\,.
\end{align}
This $\varepsilon^2$ suppression in the two-body decay rates leads to,
\begin{align}
    R_0,R_{\pm} \gtrsim 10^2\,,
\end{align}
which mirrors the same scaling as in the SM,
\begin{align}
    \frac{\text{Br}[K_L \to 3\pi^0 ]}{\text{Br}[K_L \to 2\pi^0] } \approx 200 \sim \frac{10^{-3}}{|\varepsilon|^2}\,.
\end{align}
\subsection{\texorpdfstring{$\cot \theta_\kappa \ll |\varepsilon|$}{cot theta << eps}}
\label{sec:max_FV_case_III}
We also entertain the possibility that the flavor violation in the UV is completely aligned  with $CP$ violation and the main source is coming from the SM $\varepsilon$ parameter. 
In this case the ratios are given by,
\begin{align}
    R_0(m_a) &= |\varepsilon|^2 \left(1 + \frac{1}{(1-x_a)^2}\right){I}_0\left(m_a,\left\{-1,\frac{1}{1-x_a}\right\}\right)\,,
    \\
    R_{\pm}(m_a) &=  \left(|\varepsilon|^2+\frac{1+|\varepsilon|^2}{(1-x_a)^2} \right){I}_{\pm}\left(m_a,\left\{\varepsilon\,s_{\theta_{\kappa}},\frac{-\varepsilon\,s_{\theta_{\kappa}}}{1-x_a},\frac{ s_{\theta_{\kappa}}}{1-x_a}\right\}\right).
\end{align}
The leading $CP$-conserving contribution to the neutral three-body decay rate comes from the product $\text{Im}\,\kappa^{\text{\tiny FV}} \varepsilon$, greatly suppressing the $K_L\to\pi^0 \pi^0 a$ decay rate.
For the charged decay, since the initial and final states are not  $CP$ eigenstates, the amplitude itself contains the $CP$-odd kinematic variable $\tilde{s}$ (see Eq. (\ref{Eq:kinvariables}) and discussion below).
Thus, the leading $CP$-conserving contribution to charged decay is $\text{Im}\,\kappa^{\text{\tiny FV}} \tilde{s}$ and the charged decay rate is only phase-space suppressed,
\begin{align}
    R_0(m_a=0) &= 2|\varepsilon|^2I_0(0,\{-1,1\}) \sim 10^{-9}\, ,
    \\
    R_\pm(m_a=0) &= 2I_{\pm}(0,\{0,0,1\}) \sim 10^{-4}\,.
\end{align}
\section{Minimal flavor violation}
\label{app:MFV_details}
In minimal flavor violating models, the flavor symmetry breaking in the UV theory is predominantly due to the SM Yukawas,
\begin{align}
    \text{Min}[\kappa_L^{\text{\tiny MFV}},N_8\, \mathcal{C}_i] \gg \kappa^{\text{\tiny FV}}  \,,
\end{align}
in accordance with the MFV hypothesis.
In the following we consider all the possible coupling hierarchies and their realizations from the UV perspective. In this section, we will explain the scaling found in \Sec{sec::MFV}.
%%%%%%%%%%%%%%%%%%%%%%%%%%%%%%
%%%%%%%%%%%%%%%%%%%%%%%%%%%%%%
\subsection{\texorpdfstring{$  \kappa_L^{\text{\tiny MFV}} \gg N_8\, \mathcal{C}_i $}{kappaMFV >> N8,Ci} }
%%%%%%%%%%%%%%%%%%%%%%%%%%%%%%
%%%%%%%%%%%%%%%%%%%%%%%%%%%%%%
In this case both two-body and three-body decays are dominated by $\kappa_L^{\text{\tiny MFV}} \equiv c^{\text{\tiny MFV}}V_{td}^*V_{ts}$. 
This scenario can be realized by having,
\begin{enumerate}
    \item $c^{\text{\tiny NP}}_L \neq0$, which can be realized if we allow a non-trivial ${\bm c}_{Q_L} \propto \hat{{\bm Y}}_u \hat{{\bm Y}}^\dagger_u$, consistent with MFV (see Eq. (\ref{eq:cQMFV})).
    \item $c_{tt}(\Lambda)\neq0$ .
    \item Decoupling the axion from the strong sector all together $c_{GG} = c_{Q_L} = c_{u_R}= c_{d_R}=0$, leading to $\mathcal{C}_i=0$. 
\end{enumerate}
We find identical expression for the rates as in \EqRange{eq:rates_neutral}{eq:rates_charged}, with the replacement,
\begin{align}  \theta_\kappa \to -\theta_{\text{\tiny CKM}}\,,
\end{align}
where $\theta_{\text{\tiny CKM}} \equiv \text{Arg}\,V_{td}\approx 3.5$. 
The coefficient $c^{\text{\tiny MFV}}$ is real at leading order in the MFV expansion, see the discussion in \Sec{sec:FV_couplings}. 
We thus find a sharp prediction,
\begin{align}
    R_0(m_a=0) &= 2\cot^2 \theta_{\text{\tiny CKM}}{I}_0\left(0,\left\{-1,1\right\}\right) \approx 5\times10^{-3}\,,
    \label{eq:MFV_caseI_neutral}
    \\
    R_{\pm}(m_a=0) &= \sin^{-2} \theta_{\text{\tiny CKM}}\left(1+c_{\theta_{\text{\tiny CKM}}}^2 \right){I}_\pm\left(0,\left\{-c_{\theta_{\text{\tiny CKM}}},c_{\theta_{\text{\tiny CKM}}},i\, s_{\theta_{\text{\tiny CKM}}}\right\}\right)\approx 10^{-2}\,.
\end{align}
These ratios are larger by about an order of magnitude compared to the maximally $CP$-violating case in \EqRange{eq:ratios_maximum_CP_violation_neutral}{eq:ratios_maximum_CP_violation_charged} due to a small hierarchy between the real and imaginary components of the CKM element $|\text{Re}\,V_{td}/\text{Im}\,V_{td}|\sim 2.5$. 
%%%%%%%%%%%%%%%%%%%%%%%%%%%%%%
%%%%%%%%%%%%%%%%%%%%%%%%%%%%%%
\subsection{\texorpdfstring{$N_8\,\mathcal{C}_i \gg  \kappa_L^{\text{\tiny MFV}} \gg \varepsilon\,N_8\,\mathcal{C}_i$}{N8,Ci >> kappaMFV >> eps N8}}
\label{AppendixD2}
%%%%%%%%%%%%%%%%%%%%%%%%%%%%%%
%%%%%%%%%%%%%%%%%%%%%%%%%%%%%%
In this case, the three-body decays are indirectly-mediated by the weak interaction, while the two-body decays are still dominated by $\kappa_L^{\text{\tiny MFV}}$.
The ratios are then given by,
\begin{align}
    R_{0} &= \left(\frac{|N_8|}{\text{Im}\,\kappa_L^{\text{\tiny MFV}}}\right)^2 (\mathcal{C}_2^2+\mathcal{C}_3^2)I_0\left(m_a,\left\{ \mathcal{C}_2,\mathcal{C}_3\right\}\right)\,,
\\
       R_{\pm} &= \left(\frac{|N_8|}{\text{Im}\,\kappa_L^{\text{\tiny MFV}}}\right)^2 (\mathcal{C}_4^2+\mathcal{C}_5^2+|\varepsilon|^2\mathcal{C}_6^2)I_{\pm}(m_a,\{\mathcal{C}_4,\mathcal{C}_5,\varepsilon\mathcal{C}_6\})\,.
\end{align}
For concreteness, let us assume for the remainder of this section that all the fermions couplings are flavor-blind i.e. ${\bm c}_F = c_{F}1_{3\times 3}$ for $F\in\{Q_L,u_R,d_R,L_L,e_R\}$. 
Under this assumption, $c^{\text{\tiny NP}}_L=0$.
The values of $\{c_F\}$, along with $c_{GG},c_{WW}$ and $c_{BB}$, determine the values of the couplings $\{c_{tt},\tilde{c}_{GG},\tilde{c}_{WW},\tilde{c}_{BB}\}$ as well as the combination of flavor-diagonal couplings, $\{\mathcal{C}_i\}$, appearing in the two- and three-body decay rates.

One way to realize this coupling hierarchy is by having $c_{tt}(\Lambda)=0$ and a non-vanishing $\tilde{c}_{GG}(\Lambda)$.
This can be realized by setting $c_{Q_L} = c_{u_R}$ and turning on either $c_{GG}$ or $c_{d_R}$.
In this case, the ratio of couplings varies depending on the UV scale, and, according to Eq. (\ref{eq:MFVcGG})
\begin{align}
        \left|\frac{|N_8|}{\text{Im}\,n_G V^*_{td}V_{ts}}\right|  \approx \begin{cases}
            24 \;\;\;\;\;\;\;\; &\Lambda = 10^4\,\text{GeV}
            \\
            4 & \Lambda = 10^{10}\,\text{GeV}
        \end{cases}\,.
\end{align}
All the other realizations for the hierarchy $N_8\,\mathcal{C}_i \gg  \kappa_L^{\text{\tiny MFV}} \gg \varepsilon\,N_8\,\mathcal{C}_i$ would require non-trivial cancellations to suppress the contribution from $\tilde{c}_{GG}(\Lambda)$, while still allowing for non-vanishing couplings to the strong sector such that the indirectly-mediated decay is still allowed. 
For example, by requiring that $c_{d_R} = c_{Q_L}-2c_{GG}/3$ (in addition to  $c_{Q_L}=c_{u_R}$), we set $\tilde{c}_{GG}=0$ while $\mathcal{C}_i \propto c_{GG}$.
Thus, as long as $c_{GG}\neq 0$, the indirectly-mediated processes are still accessible. 

Another possibility is realized by having $c_{Q_L}\neq0$ or $c_{L_L}\neq0$ or
$c_{WW}\neq0$. In this case, the flavor-changing coupling is dominated by the coupling to $W$ bosons with $\tilde{c}_{WW} = c_{WW}-\frac32(c_{L_L}+3c_{Q_L})$, and we find the coupling hierarchy (see Eq. (\ref{eq:MFVcWW}))
 \begin{align}
     \left|\frac{|N_8|}{\text{Im}\,n_W V^*_{td}V_{ts}}\right| \approx 40\,.
 \end{align}
 
Finally, if we set $c_{Q_L}=c_{L_L}=c_{WW}=0$, the flavor-changing coupling is dominated by the coupling to $B$ hypercharge gauge bosons with $\tilde{c}_{BB} = c_{BB}+3c_{e_R}-\frac23 c_{GG}$, and we find the coupling hierarchy (see Eq. (\ref{eq::BB_ratio}))
\begin{align}
   \left|\frac{|N_8|}{\text{Im}\,n_B V^*_{td}V_{ts}}\right|_{\Lambda = 10^{10}\,\text{GeV}}  \approx  400\,.  
\end{align}
We summarize the constraints required to realize $N_8\,\mathcal{C}_i \gg  \kappa_L^{\text{\tiny MFV}} \gg \varepsilon\,N_8\,\mathcal{C}_i$ in \Tab{tab:UV_realizations_summary}.
\begin{table}   
\begin{center}
\begin{tabular}{ |c |c| c| }
\hline
Case & Realization &   $N_8/\text{Im}\,\kappa_L^{\text{\tiny MFV}}$
\\
\hline 
$c_{tt}(\Lambda)=0 $& $ c_{u_R}-c_{Q_L}=0$ & $4-24$
\\
$\tilde{c}_{GG}(\Lambda) \neq 0$&$ c_{GG}+\frac32 (c_{d_R}-c_{Q_L})\neq0$&
\\
\hline 
$c_{tt}(\Lambda)=0 $& $ c_{u_R}-c_{Q_L}=0$ & $40$
\\
$\tilde{c}_{GG}(\Lambda) = 0$&$ c_{GG}+\frac32 (c_{d_R}-c_{Q_L})=0$&
\\
$\tilde{c}_{WW}(\Lambda)\neq 0$& $c_{WW}-\frac32 (3c_{Q_L}+c_{L_L})\neq 0$&
\\
\hline 
$c_{tt}(\Lambda)=0 $& $ c_{u_R}-c_{Q_L}=0$ & $400$
\\
$\tilde{c}_{GG}(\Lambda) = 0$&$ c_{GG}+\frac32 (c_{d_R}-c_{Q_L})=0$& (for $\Lambda = 10^{10}\,$GeV)
\\
$\tilde{c}_{WW}(\Lambda)= 0$& $c_{WW}-\frac32 (3c_{Q_L}+c_{L_L})= 0$&
\\
$\tilde{c}_{BB}(\Lambda)\neq 0$& $c_{BB} - c_{WW} - (2/3)c_{GG} + 3 c_{e_R} + 9 c_{Q_L} \neq0$&
\\
\hline
\end{tabular}
\end{center}
\caption{
Summary of the requirements on the flavor-blind UV coefficients needed to turn on the different coupling combinations $c_{tt}(\Lambda),\tilde{c}_{VV}(\Lambda)$ for $V=\{B,W,G\}$. 
Each row leads to a different coupling hierarchy, given in the last column. All these scenarios realize the hierarchy $N_8\,\mathcal{C}_i \gg  \kappa_L^{\text{\tiny MFV}} \gg \varepsilon\,N_8\,\mathcal{C}_i$ (see App. \ref{AppendixD2}).}
\label{tab:UV_realizations_summary}
\end{table}
We conclude that in this scenario, 
\begin{align}
    R_{0},R_{\pm} \sim 10^{-1}-10^{3}\,,
\end{align}
depending on the coupling hierarchy.
%%%%%%%%%%%%%%%%%%%%%%%%
%%%%%%%%%%%%%%%%%%%%%%%%
\subsection{\texorpdfstring{$\varepsilon
\, N_8\,\mathcal{C}_i \gg \kappa_L^{\text{\tiny MFV}} $}{eps N8 >> kappaMFV}}
%%%%%%%%%%%%%%%%%%%%%%%%
%%%%%%%%%%%%%%%%%%%%%%%%
In this case, both the two- and three-body decays are indirectly mediated by the weak interaction. 
The two-body decay, however, requires $N_8$ and $\varepsilon$ to break both the flavor and the $CP$ symmetries, and this suppresses the process compared to the CP-preserving three-body decay. 
The ratios are given by,
\begin{align}
    R_{0} &= \frac{1}{|\varepsilon|^2}\frac{\mathcal{C}_2^2+\mathcal{C}_3^2}{\mathcal{C}^2_1}I_0\left(m_a,\left\{ \mathcal{C}_2,\mathcal{C}_3\right\}\right)\,,
\\
       R_{\pm} &= \frac{1}{|\varepsilon|^2} \frac{\mathcal{C}_4^2+\mathcal{C}_5^2+|\varepsilon|^2\mathcal{C}_6^2}{\mathcal{C}_1^2}I_{\pm}(m_a,\{\mathcal{C}_4,\mathcal{C}_5,\varepsilon\mathcal{C}_6\})\,.
\end{align}
This scenario is realized for low UV scales $\Lambda \ll 10^6\,$GeV if only $\tilde{c}_{BB}(\Lambda)$ is non-vanishing.
In this case, we find
\begin{align}
    R_0,R_\pm \gtrsim 10^2\,.
\end{align}
%%%%%%%%%%%%%%%%%%%%%%%%
%%%%%%%%%%%%%%%%%%%%%%%%
\section{\texorpdfstring{Full expressions for  $\{\mathcal{C}_i\}$}{Full expressions for Ci}}
\label{app:Ci}
%%%%%%%%%%%%%%%%%%%%%%%%
%%%%%%%%%%%%%%%%%%%%%%%%
In this appendix we provide the full expressions for the $\{\mathcal{C}_i\}$ coefficients appearing in the main text. 
We express them in terms of the low-energy couplings,
\begin{align}
    {\bm k}_q = \begin{pmatrix}
    k_u & 0 & 0\\
     0& k_d & 
     \kappa_R
     \\
     0& 
          \kappa^\dagger_R
     & k_s 
    \end{pmatrix}\,, \;\;
        {\bm k}_Q = \begin{pmatrix}
    k_U & 0 & 0\\
     0& k_D & 
          \kappa_L
     \\
     0& 
          \kappa^\dagger_L
     & k_S 
    \end{pmatrix}\,.
\end{align}
The $\mathcal{C}_i$'s depend on $c_{GG}$ and the axial combination of couplings, which we denote by 
\begin{align}
        c_{q} &\equiv k_{q}-k_{Q}\,.
\end{align}
In the absence of the weak interactions ($N_8\to0$) and off-diagonal ALP coupling to quarks ($\kappa_L=\kappa_R=0$), the vector couplings are non-physical and can be removed from the theory using the field redefinition of \Eq{eq:generic_basis}.\footnote{Additional couplings would be physical in the case of a generic scalar that is not a Pseudo-Nambu-Goldstone boson, see e.g~\cite{Delaunay:2025lhl}.}
Otherwise, one linear combination of vector couplings cannot be rotated away and is therefore physical~\cite{Bauer:2021wjo}, 
\begin{align}
    \Delta k_{V}^{d-s}&\equiv k_{d}+k_{D}-(k_{s}+k_{S})\,.
\end{align}
The coefficients are given by,
\begin{align}
    \mathcal{C}_1 = &-\frac{ m_a^2 (c_d-2 c_s+c_u)}{3 m_a^2-4 m_K^2+m_{\pi}^2}+\frac{ m_a^2 (c_s-c_d)}{2 (m^2_K-m^2_{\pi})}+\frac{ m^2_a (c_d-c_u)}{m^2_a-m^2_{\pi}}-\frac32 c_d-\frac12  c_s
    \\
    &+\frac{8 (m^2_K-m^2_a) }{3 m_a^2-4 m_K^2+m_\pi^2}c_{GG}
    -\frac{m_K^2+m_\pi^2-m_a^2}{2 \left(m_K^2-m_\pi^2\right)}\Delta k_{V}^{d-s} \nonumber\,.
\end{align}
\begin{align}\label{eq:DC1}
    \Delta \mathcal{C}_1 = \frac{m_\pi^2 (c_d-c_u)}{(m^2_\pi-m_a^2) }\,.
\end{align}
\begin{align}
    \mathcal{C}_2 =& -\frac{m_a^2 (c_d-2 c_s+c_u)}{3 m_a^2-4 m_K^2+m_\pi^2}+\frac{m_a^2 (c_s-c_d)}{2 m_K^2}+\frac{m_K^2 (c_d-c_s)}{m_K^2-m_\pi^2}+\frac{m_a^2 (c_d-c_u)}{m_a^2-m_\pi^2}-\frac{5 c_d}{2}+\frac12 c_s
    \\
    &+\frac{8 (m^2_K-m^2_a) }{3 m_a^2-4 m_K^2+m_\pi^2}c_{GG} +
  \frac{m_a^2-m_K^2+4 m_\pi^2}{2 m_K^2}\Delta k_{V}^{d-s}\nonumber\,.
\end{align}
\begin{align}
    \mathcal{C}_3 =& \frac{m_\pi^2 \left(3 m_a^2 (c_d-c_s)+m_K^2 (c_s-5 c_d)\right)+3 m_a^2 m_K^2 (c_s-c_d)+m_K^4 (3 c_d+c_s)}{2  \left(m_K^2-m_a^2\right) (m_K^2-m^2_\pi) }-\frac32\Delta k_{V}^{d-s}\,.
\end{align}
\begin{align}
  \mathcal{C}_4 =&  -\frac{ m_a^2 (c_d-2 c_s+c_u)}{3 m_a^2-4 m_K^2+m_\pi^2}+\frac{m_a^2 (c_d+c_s-2 c_u)}{2m_K^2}+\frac{m^2_K (c_d-c_s)}{m^2_K-m^2_\pi}+\frac{m_a^4 (c_u-c_d)}{m_K^2 (m^2_a-m^2_\pi)}\nonumber
  \\
  & -\frac32 c_d+\frac12 c_s- c_u
  +\frac{8 (m^2_K-m^2_a) }{3 m_a^2-4 m_K^2+m_{\pi}^2}c_{GG}
  +
  \frac{m_a^2-m_K^2+4 m_\pi^2}{2 m_K^2}\Delta k_{V}^{d-s}\,.
\end{align}
\begin{align}
    \mathcal{C}_5 =& \frac{m_\pi^2}{m_\pi^2-m_a^2} c_u 
    -\frac{3 m_a^2 (m_K^2-m^2_\pi) +m_K^2 \left(m_K^2+m_\pi^2\right)}{2 (m^2_a-m^2_K) (m^2_K-m^2_\pi)}c_s
    \nonumber
    \\
    &+
    \frac{3 m_a^4 (m_K^2-m^2_\pi)+m_a^2 \left(-3 m_K^4+4 m_K^2 m_\pi^2+m_\pi^4\right)+m_K^2 m_\pi^2 \left(m_K^2-3 m_\pi^2\right)}{2 (m^2_a-m^2_K) (m_a^2-m^2_\pi)  (m_K^2-m^2_\pi) } c_d\,.
\end{align}
\begin{align}
    \mathcal{C}_6 = &  -\frac{3 m_a^2 (c_d-2 c_s+c_u)}{3 m_a^2-4 m_K^2+m_\pi^2}+\frac{m_a^2 (c_d-3 c_s+2 c_u)+m_K^2 (-3 c_d+c_s-2 c_u)}{2(m^2_a-m^2_K) }
    \\
     & + \frac{m_K^4 (c_d-c_s)}{(m^2_a-m^2_K)  (m^2_K-m^2_\pi)}-\frac{8 (m^2_K-m^2_\pi) }{3 m_a^2-4 m_K^2+m_\pi^2}c_{GG}+\frac12 \Delta k_{V}^{d-s}\nonumber\,.
\end{align}
\begin{align}
    \mathcal{C}_{27} =& -\frac{4 (m_a^2-m^2_K) }{3 m_a^2-4 m_K^2+m_{\pi}^2} c_{GG}+\frac{m_a^2-m_K^2-m_{\pi}^2}{2 \left(m_K^2-m_{\pi}^2\right)}\Delta k_{V}^{d-s}\\
    &+\frac{1}{2} \left(3 c_d+c_s-4 c_u\right) 
    +\frac{3 m_a^2 \left(c_u-c_d\right)}{2 (m_a^2-m_{\pi}^2) }+\frac{m_a^2 \left(c_s-c_d\right)}{2 (m_K^2-m_{\pi}^2) }-\frac{m_a^2 \left(c_d-2 c_s+c_u\right)}{2 \left(3 m_a^2-4 m_K^2+m_{\pi}^2\right)}\,.\nonumber
\end{align}
For light ALP masses $m^2_a \ll m^2_\pi \ll m^2_K$,
\begin{align}
    \mathcal{C}_1 \to &\frac{1}{2} (-4 c_{GG}-3 c_d-c_s-\Delta k_{V}^{d-s})\,,
    \\
      \mathcal{C}_2 \to &\frac{1}{2} (-4 c_{GG}-3 c_d-c_s-\Delta k_{V}^{d-s})\,,
          \\
      \mathcal{C}_3 \to &\frac{1}{2} (3 c_d+c_s-3 \Delta k_{V}^{d-s})\,,
          \\
      \mathcal{C}_4 \to &\frac{1}{2} (-4 c_{GG}-c_d-c_s-2 c_u-\Delta k_{V}^{d-s})\,,
          \\
      \mathcal{C}_5 \to &\frac{1}{2} (c_d+c_s+2 c_u-3 \Delta k_{V}^{d-s})\,,
          \\
      \mathcal{C}_6 \to &\frac{1}{2} (4 c_{GG}+c_d+c_s+2 c_u+\Delta k_{V}^{d-s})\,,
      \\
      \mathcal{C}_{27} \to& \frac{1}{2} (3 c_d+c_s-4 c_u-2 c_{GG}-\Delta k_{V}^{d-s})\,.
\end{align}
All the expressions above are given in terms of the low-energy couplings, which in principle could differ from the UV coupling due to the running. In the main text we report the expression for the $ \mathcal{C}_i$'s in the $m_K \gg m_\pi \gg m_a$ approximation in the limit where the RGE running can be neglected, expressed in term of the UV couplings.
\section{ALP coupling to photons}
\label{app:photon_coupling}
The effective and basis-independent low-energy ALP coupling to photons is given by~\cite{Bauer:2020jbp,Bauer:2017ris},
\begin{align}
C_{\gamma\gamma}^{\rm eff}
=& c_{\gamma\gamma}
- \left( \frac{5}{3}
    + \frac{m_\pi^2}{m_\pi^2 - m_a^2}
      \frac{m_d - m_u}{m_u + m_d}
  \right) c_{GG}
- \frac{m_a^2}{m_\pi^2 - m_a^2}
  \frac{c_{uu}(\mu_\chi) - c_{dd}(\mu_\chi)}{2}\, \nonumber
  \\
  &
  +\sum_{\ell=e,\mu} c_{\ell\ell}(\mu_\chi)B_1\left(\frac{4 m_\ell^2}{m_a^2} \right)
  \label{eq::Cgamma_eff}
\end{align}
where $\mu_\chi\sim\text{GeV}$.
The above combination of low-energy couplings to light quarks can be written in terms of the UV couplings as~\cite{Bauer:2020jbp},
\begin{align}
    c_{uu}(\mu_\chi) - c_{dd}(\mu_\chi)
\simeq
c_{uu}(\Lambda) - c_{dd}(\Lambda)
- 6\, \frac{\alpha_t(m_t)}{\alpha_s(m_t)}
\left[ 1 - \left( \frac{\alpha_s(\Lambda)}{\alpha_s(m_t)} \right)^{1/7} \right]
c_{tt}(\Lambda)\, ,
\end{align}
with $c_{qq}(\Lambda) \equiv \big[{\bm c}_{q_R}(\Lambda)\big]_{11}-\big[\cQL(\Lambda)\big]_{11}$ for $q=\{u,d\}$ and $c_{tt}(\Lambda)\equiv\big[{\bm c}_{u_R}(\Lambda)\big]_{33}-\big[\cQL(\Lambda)\big]_{33}$.
In the last term of \Eq{eq::Cgamma_eff} we included the loop contributions from the lightest charged leptons, with
\begin{align}
    B_1(\tau)\equiv 1-\tau f^2(\tau)\, \;\;\;\; f(\tau)=\begin{cases}
        \arcsin(1/\sqrt{\tau}) \;\;\;\;\; &\tau \geq 1
        \\
        \frac{\pi}{2}+\frac{i}{2} \log \frac{1+\sqrt{1-\tau}}{1-\sqrt{1-\tau}} & \tau<1
    \end{cases}\,.
\end{align}
The low-energy axial coupling to leptons is given by,
\begin{align}
c_{\ell\ell}(\mu_\chi) = c_{\ell\ell}(\Lambda)+\Delta c\,,
\end{align}
for $\ell=\{e,\mu\}$. 
The UV axial coupling of the electron is defined as
$c_{ee}(\Lambda) \equiv [c_{e_R}(\Lambda)]_{11}-[c_{L_L}(\Lambda)]_{11}$, and similarly for the muon.
The universal running contribution is given by~\cite{Bauer:2020jbp}
\begin{align}
    \Delta c = &6n_{t}c_{tt}(\Lambda)+6n_{G}\tilde{c}_{GG}(\Lambda)\nonumber
    \\
    &+\tilde{c}_{WW}(\Lambda) \left[\frac{9}{16\pi^2}\int_\Lambda^{\muEW}\frac{d\mu}{\mu}
    \left[6n_t(\muEW,\mu)+1\right]
     \alpha_2^2(\mu)\right]\nonumber
    \\
    &+\tilde{c}_{BB}(\Lambda) \left[\frac{15}{16\pi^2}\int_\Lambda^{\muEW}\frac{d\mu}{\mu}
    \left[(34/15)n_t(\muEW,\mu)+1\right]
     \alpha_1^2(\mu)\right]\,.
\end{align}
See \Eq{eq:nt_def_1} and \Eq{eq:nG}  for the definitions of $n_t$ and $n_G$, respectively. 
\section{Experimental recasts}
In this appendix we provide some details about the recasts of the experimental results used for  \Fig{fig:K_bounds}. 
The recasts were needed in cases where the experimental results were given for a stable ALP i.e. in the limit of infinite lifetime. For each experiment, we estimate the experimental efficiency as a function the finite lifetime of the ALP.
The ALP lifetime in the model we consider in \Sec{section:prompt_ALP} is effectivity $\tau_a \approx \Gamma^{-1}(a\to \gamma \gamma)$. 
\label{app:exp_recasts}
\subsection{\texorpdfstring{KOTO $K_L \to \pi^0 X$}{KOTO KL to pi0 X}}
\begin{figure}[t]
  \centering   \includegraphics[width=0.5\textwidth]{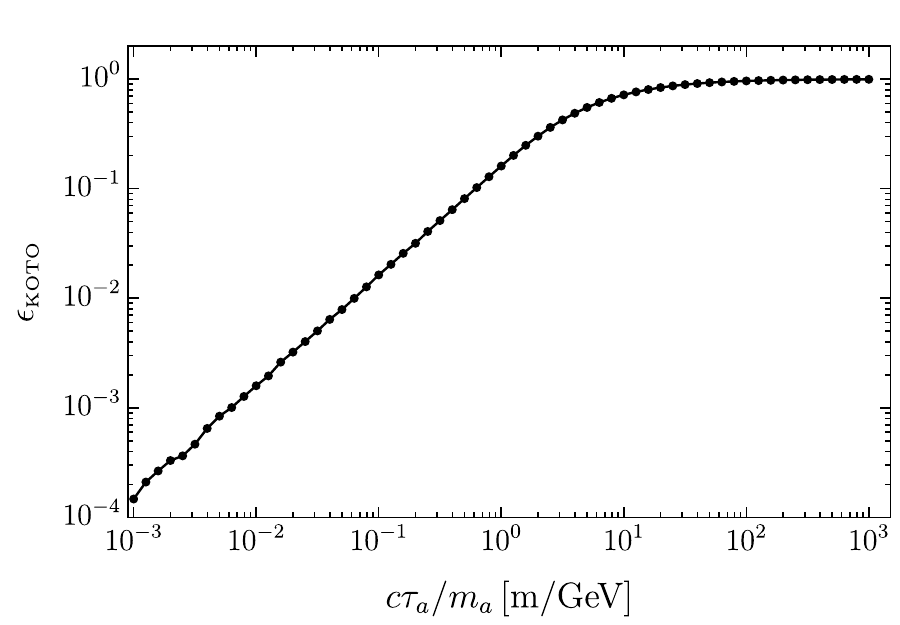}  
      \caption{Estimated signal efficiency for $K_L \to \pi^0 X$ in KOTO as a function of the combination  $c\tau_a/m_a$, see \Eq{eq:KOTO_Eff}. }     \label{plot::koto_eff}
\end{figure}

We estimate the signal efficiency for the KOTO search $K_L \to \pi^0 X$~\cite{KOTO:2024zbl} by
\begin{align}
\label{eq:KOTO_Eff}
   \epsilon_{\text{\tiny KOTO}}(c\tau_a,m_a) =  \int_0^{L_{\text{\tiny KOTO}}} \frac{\mathrm{d}z_{K_L}}{L_{\text{\tiny KOTO}}} \int_{-1}^{+1}\frac{\mathrm{d}\cos \theta_a }{2}\int_0^{\infty} \frac{\mathrm{d}\ell_a}{\bar\ell_a}e^{-\ell_a/\bar{\ell}_a}\mathcal{A}(\ell_a,\theta^{\text{\tiny lab}}_a,z_{K_L})
\end{align}
where 
\begin{align}
\label{eq::KOTO_defs}
    \bar\ell_a(\theta_a,c\tau,m_a) &\equiv\frac{c\tau }{m_a}|\vec{p}_a(p_K,\theta_a)|\,,
    \\
    \label{eq:pa_lab}
    \vec{p}_a(p_K,\theta_a) &= \left(p_{\text{\tiny cm}}\sin\theta_a,0,\sqrt{1+\frac{p_K}{m_K}}p_{\text{\tiny cm}}\cos \theta_a+ \frac{p_K}{m_K}p_{\text{\tiny cm}}\right)\,,
    \\\label{eq:pcm}
    p_{\text{\tiny cm}} &= \frac{\sqrt{(m_K^2-(m_\pi-m_a)^2)(m_K^2-(m_\pi+m_a)^2)}}{2m_K},
    \\
    \label{eq:KOTO_acc}
    \mathcal{A}(\ell_a,\theta^{\text{\tiny lab}}_a,z_{K_L}) &= \begin{cases}
        0 \;\;\;\;\;& z_{K_L}+\ell_a \cos \theta^{\text{\tiny lab}}_a < L_{\text{\tiny KOTO}} \;\;\;\;\text{and}\;\;\;\; \ell_a \sin \theta^{\text{\tiny lab}}_a <R_{\text{\tiny KOTO}}
        \\
        1 & \text{otherwise}\,,
    \end{cases}
\end{align}
where $\theta_a^{\text{\tiny lab}}$ is the ALP angle in the lab frame implied by \Eq{eq:pa_lab}, not to be confused with $\theta_a$ defined in the rest frame of the decaying particle.
We approximate the KOTO detector as a cylinder of radius $ R_{\text{\tiny KOTO}}=1\,\text{m}$ and length $L_{\text{\tiny KOTO}}=4.148\,\text{m}$, 
with \Eq{eq:KOTO_acc} requiring the ALP decays outside that cylinder.
We further assume the kaon beam to be approximately monochromatic with $p_K = p_{\text{\tiny KOTO}} = 1.5\,\text{GeV}$, which corresponds to the average momentum of kaons in the beam. 
Due to its large decay length in the lab frame ($\approx 10\,\text{m}$), we assume that the kaon decay vertex is distributed uniformly inside the cylinder.
The resulting efficiency depends strongly only on the combinations $c\tau_a/m_a$, with mass dependence becoming important only close to the threshold $m_a \approx m_K-m_\pi$. 
We plot the efficiency in \Fig{plot::koto_eff}.
\subsection{\texorpdfstring{E949 $K^+ \to \pi^+ \gamma \gamma$}{E949 K+ to pi+ gamma gamma}}
We estimate the signal efficiency for the E949 $K^+\to \pi^+ \gamma \gamma$ search~\cite{E949:2005qiy} by,
\begin{align}
    \epsilon_{\text{\tiny E939}}(c\tau_a,m_a)= \int_0^{R_{\text{E949}}}\frac{\mathrm{d} \ell_a}{\bar\ell_{a,0}} e^{-\ell_a/\bar\ell_{a,0}} = 1-e^{-R_{\text{\tiny E949}}/\bar\ell_{a,0}}\,,
\end{align}
where in this case since the kaons decay effectivity at rest,
\begin{align}
   \bar \ell_{a,0}\equiv\bar\ell_a(p_K=0,\theta_a,c\tau,m_a) = \frac{c \tau_a p_{\text{\tiny cm}}}{m_a},
\end{align}
and $p_{\text{\tiny cm}}$ is defined in \Eq{eq:pcm} and we used $R_{\text{\tiny E949}} = 1.45\,$m.
\subsection{\texorpdfstring{NA48 $K_L \to \pi^0 \gamma \gamma$}{NA48 KL to pi0 gamma gamma}}
In order to estimate the signal efficiency for the NA48 $K_L \to \pi^0 \gamma \gamma$ search, we simulated $10^6$ $K_L$ decay events for 5 ALP masses $m_a =\{0.01,0.1,0.21,0.29,0.35\}\,$GeV and 41 $c\tau_a$ values, logarithmically distributed in the range $[10^{-4}\,\text{m},10^{4}\,\text{m}]\,$. 
In each simulation point we decay a $K_L$ to $\pi^0 a$ in an arbitrary position inside the decay volume, $z_{K_L}\in[Z_{\text{\tiny cal,start}},Z_{\text{\tiny cal,end}}] = [126\,\text{m},156\,\text{m}]$.
The kaon carries a momentum $p_K$ which is drawn according to the $K_L$ momentum distribution in NA48, see Fig. (24) in Ref.~\cite{NA48:2001bct}.
We propagate the ALP a distance $\ell_a$ drawn from the appropriate exponential distribution. 
The propagation distance depends on the ALP momentum, which is drawn with an arbitrary angle $\theta_a$ in the COM frame and then boosted to the lab frame according to $p_K$.
If the ALP propagates beyond the position of the calorimeter (ECAL), namely if
\begin{align}\label{eq:geomacc}
    z_{K_L}+\ell_a \cos \theta_a^{\text{\tiny COM}} > Z_{\text{\tiny ECAL}} = 241\,\text{m}  \;\;\;\; (\text{geometric})\,,
\end{align}
we assign efficiency $0$. 
This geometrical acceptance criterion provides the strongest dependence on the ALP lifetime and the main source for efficiency loss.
Other selection criteria were used in this search in order to reduce backgrounds. We find that the only selection criterion which had non-negligible effect on the signal efficiency for a finite-lifetime ALP is the requirement that the $K_L$ originates from the decay volume. The $K_L$ decay vertex is deduced from the reconstructed decay vertex,
\begin{align}
    z_{K_L}^{\text{\tiny recon.}} \equiv Z_{\text{\tiny ECAL}}- \frac{\sum_{i>j} E_{\gamma, i}E_{\gamma,j}d^2_{ij}}{m_K}\,,
\end{align}
where $E_{\gamma,i}$ is the $i$-th photon energy and $d_{ij}$ is the transverse distance between the $i$-th and $j$-th photons on the ECAL plane.
The sum runs for all non-identical photon pairs without repetitions with $i,j=\{1,..,4\}$.
The selection criteria is defined as,
\begin{align}
    z_{K_L}^{\text{\tiny recon.}} \in[Z_{\text{\tiny cal,start}},Z_{\text{\tiny cal,end}}] = [126\,\text{m},156\,\text{m}] \;\;\;\; (\text{selection 1})\,.
\end{align}
In the limit where all the photons originate from the same vertex, this formula is reliable for small angles, which is a very good approximation due to the large kaon boost. 
However, when the ALP is significantly displaced, the reconstructed $K_L$ vertex position gets biased towards larger values, leading to a reduction in signal efficiency when $z_{K_L}^{\text{\tiny recon.}}$ is pushed outside the decay volume.

In addition, we require that the two photons originating from the ALP do not merge and can be reconstructed, since otherwise the events would only register 3 photons in the final state and would not be recorded. 
We enforce this criterion by requiring that
\begin{align}
    d_{34} > \sigma_{\text{\tiny ECAL}}\;\;\;\;(\text{selection 2})\,,
\end{align}
where we used $\sigma_{\text{\tiny ECAL}}= 1.3\times 10^{-3}\,\text{m}$~\cite{NA48:2001bct} and $d_{34}$ is the transverse distance between the two photons originating from the ALP.
\begin{figure}[t]
  \centering   \includegraphics[width=0.7\textwidth]{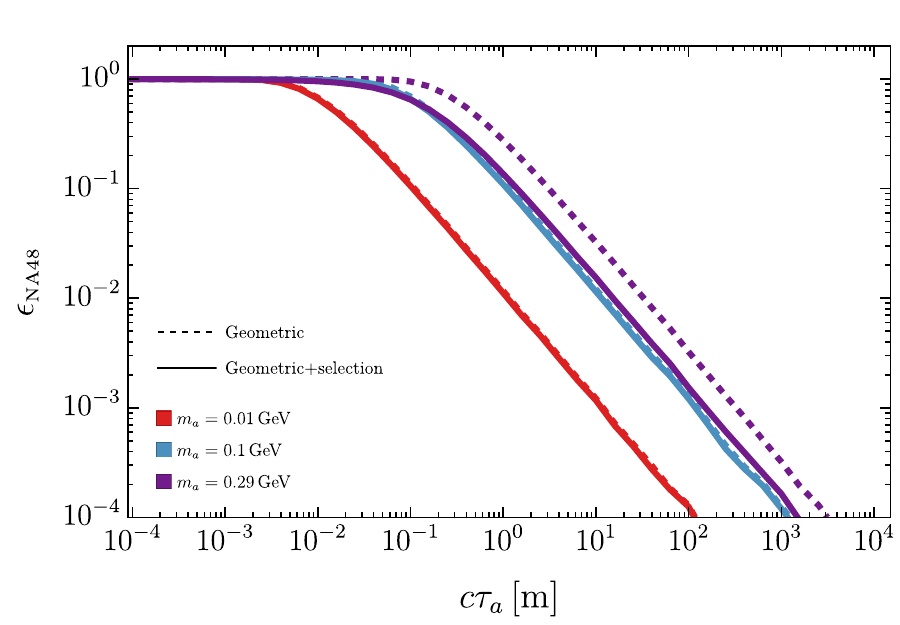}  
      \caption{Estimate of the signal efficiency for the NA48 $K_L \to \pi^0 \gamma \gamma$ search as a function of $c\tau_a$ for $m_a=0.01,0.1$ and $0.29\,\text{GeV}$ in red, blue and purple, respectively. The solid lines represent the full efficiency, while the dashed lines are the efficiency if only the geometric acceptance in Eq. (\ref{eq:geomacc}) is taken into account.}     \label{plot::NA48_eff}
\end{figure}
We plot the resulting efficiencies in \Fig{plot::NA48_eff} for three selected masses as a function of $c\tau_a$. 
The solid lines represent the full efficiencies, and the dashed lines show the efficiencies using only the geometric acceptance criterion in Eq. (\ref{eq:geomacc}). 
At lower masses, the selection criteria have a negligible effect and the full efficiency is only due to geometrical acceptance and is essentially a function of $c\tau_a/m_a$, as expected.
At higher masses, the selection criteria lead to an $\mathcal{O}(1-10)$ reduction in efficiency.
Lastly, we take advantage of the fact that the bounds in Ref.~\cite{NA48:2001bct} are given in bins of $m_{34} \equiv \sqrt{(p_{\gamma_3}+p_{\gamma_4})^2}$, which in our case would just be the ALP mass $m_a$, namely for a given ALP mass the signal would contribute to a single bin.  
For a given ALP mass, we then use the upper limit of the branching fraction for the relevant bin, weighted with the appropriate   
efficiency factor (interpolated if needed) to set the limits shown in \Fig{fig:K_bounds}. 
\subsection{\texorpdfstring{E391a $K_L \to \pi^0 \pi^0 X$}{E391a KL to pi0 pi0 X}}
We estimate the signal efficiency in the E391a $K_L\to \pi^0\pi^0X$ search~\cite{E949:2005qiy} for an invisible particle $X$, by,
\begin{align}
    \epsilon_{\text{\tiny E391a}}(c\tau_a,m_a)= \int_{R_{\text{E391a}}}^{\infty}\frac{\mathrm{d} \ell_a}{\bar\ell^{\text{\tiny E391a}}_{a}} e^{-\ell_a/\bar\ell^{\text{\tiny E391a}}_{a}} = e^{-R_{\text{\tiny E391a}}/\bar\ell^{\text{\tiny E391a}}_{a}}\,,
\end{align}
where
\begin{align}
    \bar\ell^{\text{\tiny E391a}}_{a}(c\tau_a,m_a) \equiv \frac{c \tau_a p_a^{\text{\tiny E391a}}}{m_a}\,.
\end{align}
The typical kaon momentum in E391a was $p^{\text{\tiny E391a}}_{K_L}\approx 1.5\,$GeV, thus we estimate that the typical ALP   produced via the three-body decay has momentum $p_a^{\text{\tiny E391a}}\approx p^{\text{\tiny E391a}}_{K_L}/3 \approx 0.5\,$GeV, and we used $R_{\text{\tiny E391a}} = 2\,$m.
\subsection{\texorpdfstring{BABAR $\Upsilon(1S)\to \gamma X$}{BaBaR Upsilon(1S)-> gamma X}}
The Babar search in~\cite{BaBar:2010eww} targeted the decay
\begin{align}
    \Upsilon(1S) \to \gamma\, X\,,
\end{align} 
with $X$ escaping detection. The $\Upsilon(1S)$ particles were produced from the decay 
$\Upsilon(2S) \to \Upsilon(1S) \pi^+ \pi^-$. 
On the $\Upsilon(2S)$ pole the Babar beam energies were set to be $E_{e^-}= 8.05\,$GeV and $E_{e^+}=3.12\,$GeV, producing a slightly boosted $\Upsilon(2S)$.
We simulate the produced ALP spectrum by first decaying the $\Upsilon(2S)$ assuming a uniform distribution in the three-body Dalitz plane, followed by the two-body decay $\Upsilon(1S) \to \gamma\, a$.
We conservatively require the ALP to decay outside the box-shaped region containing the electromagnetic calorimeter, where $\Delta y = 1.375\,$m, $\Delta z_F=2.295\,$m and $\Delta z_B=1.555\,$m~\cite{BaBar:2001yhh}  are the distances between the interaction point to the top, front and back of the box, respectively.
We calculate the efficiency numerically for various values of $c\tau_a/m_a$.
The resulting efficiency is  well-approximated by,
\begin{align}
    \epsilon_{\text{\tiny BABAR}} (c\tau_a/m_a) =e^{ -{R_{\text{\tiny BABAR}}}/{  \bar{\ell}_a^{\text{\tiny BABAR}}}}\,,
\end{align}
where our fit value $R_{\text{\tiny BABAR}} = 1.8\,$m is consistent with the average distance the ALP needs to propagate to escape detection, and
\begin{align}    \bar{\ell}_a^{\text{\tiny BABAR}} \equiv p_a^{\text{\tiny BABAR}}c\tau_a/m_a\,.
\end{align}
Our fitted value $ p_a^{\text{\tiny BABAR}}=5.27\,$GeV is consistent with the average momentum of the ALP.

% Bibliography

\bibliographystyle{JHEP}
\bibliography{biblio.bib}

\end{document}